\documentclass[%
reprint,
notitlepage,
 showpacs,preprintnumbers,
 aps,
pra,
 longbibliography
]{revtex4-1}

\usepackage{amsmath}
\usepackage{amsthm}
\usepackage{dsfont}
\usepackage{mathbbol}
\usepackage{bbold}
\usepackage{mathrsfs}
\usepackage{graphicx}
\usepackage{dcolumn}
\usepackage{bm}
\usepackage{comment}
\usepackage{verbatim}
\usepackage[usenames]{color}
\usepackage[dvipsnames]{xcolor}
\usepackage[normalem]{ulem}

\DeclareMathOperator*{\SumInt}{%
\mathchoice%
  {\ooalign{$\displaystyle\sum$\cr\hidewidth$\displaystyle\int$\hidewidth\cr}}
  {\ooalign{\raisebox{.14\height}{\scalebox{.7}{$\textstyle\sum$}}\cr\hidewidth$\textstyle\int$\hidewidth\cr}}
  {\ooalign{\raisebox{.2\height}{\scalebox{.6}{$\scriptstyle\sum$}}\cr$\scriptstyle\int$\cr}}
  {\ooalign{\raisebox{.2\height}{\scalebox{.6}{$\scriptstyle\sum$}}\cr$\scriptstyle\int$\cr}}
}

\newcommand{\wf}{wave function}
\newcommand{\triplets}{\ensuremath{{}^3\text{S}}}
\newcommand{\doublets}{\ensuremath{{}^2\text{S}}}
\newcommand{\singlets}{\ensuremath{{}^1\text{S}}}
\newcommand{\quintetsigma}{\ensuremath{{}^5\Sigma}_\text{g}}
\newcommand{\tripletsigma}{\ensuremath{{}^3\Sigma}_\text{u}}
\newcommand{\singletsigma}{\ensuremath{{}^1\Sigma}_\text{g}}
\newcommand{\doubletsigmag}{\ensuremath{{}^2\Sigma}_\text{g}}
\newcommand{\doubletsigmau}{\ensuremath{{}^2\Sigma}_\text{u}}

\usepackage[per-mode=symbol]{siunitx}
\usepackage[version=3]{mhchem}

\DeclareSIUnit\intensity{\watt\per\centi\meter\squared}
\DeclareSIUnit\fieldstrength{\volt\per\centi\meter}
\DeclareSIUnit\kfieldstrength{k\volt\per\centi\meter}
\DeclareSIUnit\energy{cm^{-1}}

\newcommand{\melement}[3]{\ensuremath{\left\langle #1 \left|#2\right|#3\right\rangle}}%
\newcommand{\ie}{i.\,e.}%
\newcommand{\ket}[1]{\left|#1\right\rangle}

\newcommand{\uoft}{\affiliation{Chemical Physics Theory Group, Department of Chemistry, and Center for Quantum Information and Quantum Control, University of Toronto, Toronto, ON M5S 3H6, Canada}}

\newcommand{\uam}{\affiliation{Departamento de Qu\'imica, M\'odulo 13, Universidad Aut\'onoma de Madrid, Madrid, 28049, Spain}}
\newcommand{\physreno}{\affiliation{Department of Physics, University of Nevada, Reno, NV, 89557, USA}}

\begin{document}
\title{Coherent control mechanisms of Penning and associative ionization in cold He$^*({2}^3\text{S})$-He$^*({2}^3\text{S})$ reactive scattering}
\author{Juan J. Omiste}\email{juan.omiste@uam.es}\uam
\author{Timur V. Tscherbul}\physreno
\author{Paul Brumer}\uoft
\date{\today}
\begin{abstract}
We explore
coherent control of Penning and associative ionization in cold  collisions of metastable He$^*({2}^3\text{S})$ atoms via the quantum interference between different states of the He$_2^*$ collision complex. By tuning the preparation coefficients of the initial atomic spin states, we can benefit from the quantum interference between molecular channels to maximize or minimize the cross sections for Penning and associative ionization. {In particular, we find that we can enhance the ionization ratio by 30\% in the cold regime.} This work is significant for the coherent control of chemical reactions in the cold and ultracold regime.
\end{abstract}

\maketitle
\newpage

\section{Introduction}
\label{sec:introduction}
Developing new tools and techniques for controlling cold and ultracold atomic collisions has been the focus of numerous experimental and theoretical studies~\cite{Chin:10,Vassen:12} due to the pivotal role of interatomic interactions in determining the collective properties of ultracold atomic gases and Bose-Einstein condensates~\cite{Chin:10,Vassen:12}.  For example, the use of Feshbach resonances induced by external magnetic or laser fields enables precise tuning of interatomic interactions in optical lattices and the realization of strongly interacting  quantum states of matter \cite{Bloch:08,Bloch:12}. Apart from magnetic Feshbach resonances, a number of alternative  mechanisms have been proposed for controlling  ultracold atomic collisions based on dc and ac electric fields~\cite{Krems:06,Devolder:19}, radiofrequency fields~\cite{Tscherbul:10,Hanna:10,Ding:17} and laser radiation~\cite{Fedichev:96,Bohn:97,Nicholson:15}.  

  Coherent control  \cite{Shapiro2012} is a promising technique for manipulating ultracold atomic   \cite{Omiste2018_penning}  and molecular  \cite{Devolder:20} scattering dynamics by initiating scattering  in a quantum superposition of internal states, \cite{Brumer1996,Abrashkevich2001} which leads to constructive and destructive interference between  the different indistinguishable  scattering pathways, affecting the outcome of the dynamical process \cite{Shapiro2012}. Initially applied to laser-driven unimolecular processes such as photodissociation, coherent control has enjoyed much success when applied to a wide range of molecular processes.
 Due to the small number of available quantum states and their robustness to decoherence at low temperatures, cold collisions could be particularly amenable to coherent control,  providing a fertile ground for developing and applying new control scenarios.  Recent theoretical and experimental studies demonstrated efficient quantum state control of product channel branching in Penning ionization (PI) and associative ionization (AI) in cold collisions of ground-state Ar atoms with metastable Ne$^*$  \cite{Gordon2018}. Additional theoretical studies have explored  coherent control of cold Ar~+~Ne$^*$  collisions, stressing the important role of rotational symmetry  \cite{Arango2006a,Arango2006b,Omiste2018_penning}. (See, however, Ref. \footnote{As noted in Ref. [13], there is a computational error in References [17] and [18].  The formalism in these papers is correct, and referenced in this paper, but the computed control results are incorrect and superceded by Ref. [13]}).

Our previous  work \cite{Arango2006a,Arango2006b,Omiste2018_penning} has focused on coherent control of cold collisions of metastable and ground-state rare-gas atoms such as Ne$^*(^3\text{P}_2)$~+~Ar, raising the question of whether low-temperature collisions of {\it two} metastable atoms such as He$^*$~+~He$^*$ or  Ne$^*$~+~Ne$^*$ can be efficiently controlled using quantum superpositions. Such collisions  play a key role in  evaporative cooling of trapped metastable rare-gas  atoms \cite{Vassen:12,Doret:09}. Specifically, while elastic collisions drive thermalization and cooling, inelastic collisions (PI and AI) lead to detrimental trap losses \cite{Vassen:12}. Thus, minimizing inelastic collision rates could be used to optimize evaporative cooling of metastable atoms, and thereby lead to denser and longer-lived ultracold atomic gases \cite{Vassen:12}.

 He$^*$ plays a particularly significant role in ultracold  atomic physics \cite{Vassen:12} as the first metastable atom cooled to quantum degeneracy \cite{PereiraDosSantos2001,Robert:01,Doret:09}, as well as the lightest and simplest of all rare gas atoms with a unique spherically symmetric $^3$S electronic configuration. These characteristics could make He$^*(\triplets)$ an ideal collision partner for  sympathetic cooling of atoms and molecules in a magnetic trap~\cite{Barletta2009,Tscherbul2011}.
In addition, interactions of He$^*$ with other atoms, such as Rb, are readily amenable to  highly accurate {\it ab initio} and quantum scattering calculations \cite{Knoop:14}.  Penning and associative ionization in cold He$^*\left(2\triplets\right)$+He$^*\left(2\triplets\right)$ collisions could therefore serve as a paradigm of collisional processes in cold mixtures of metastable rare gas atoms.

Here, we explore the possibility of coherent control over  inelastic AI and PI processes  in cold and ultracold collisions of metastable He$^*$ atoms.
  We show that by forming coherent superpositions of the degenerate magnetic sublevels of He$^*$, it is possible to effectively enhance or suppress the PI and AI cross sections in cold He$^*$~+~He$^*$  collisions, as well as the branching ratio of AI to PI.

The remainder of this paper is organized as follows. Section~\ref{sec:theory_pi_ai}  describes the general theory and numerical methods used to compute the Penning and associative ionization cross section in atom-atom scattering. The theory for He$^*\left(2\triplets\right)$+He$^*\left(2\triplets\right)$ collisions is described in Section~\ref{sec:case_of_hestar_hestar_scattering}. Computational results are provided in Sec.~\ref{sec:results} and conclusions discussed in Sec.~\ref{sec:conclusions_and_outlook}.

\section{ Penning and associative ionization}
\label{sec:theory_pi_ai}

To describe the theory of the scattering processes that lead to Penning and associative ionization, we focus on the case of two colliding atoms. However, the treatment can also be extended to atom-molecule or molecule-molecule scattering ~\cite{Siska1993,Arango2006b,Muller1991}. Details on the numerical methods used are provided in Appendix~\ref{sec:scattering_states_xs}.

An atom $A$ in the metastable state $A^*$ that collides with an atom $B$ can undergo either Penning ionization (PI)
  \begin{equation}
    \label{eq:astarb_penning}
    A^*+B\rightarrow A+B^++e^-,
  \end{equation}
  or  associative ionization (AI)
  \begin{equation}
    \label{eq:astarb_associative}
    A^*+B\rightarrow (AB)^++e^-.
  \end{equation}
The products of PI are  atom $A$ de-excited to its ground-state, an ion $B^+$ and the ejected electron $e^-$,~\i.e., the ~\emph{Penning electron}. AI leads to a dimer $AB^+$ and an ejected electron. The energy balance of the reaction is given by~\cite{Siska1993}
\begin{equation}
\label{eq:energy_balance}
  E_*+\varepsilon_0=\text{IE}+\varepsilon+E_+,
\end{equation}
where $\varepsilon_0$ is the electronic excitation energy of the atom $A^*$, $E_*$ is the incident kinetic energy of the collision, IE is the first ionization energy of $B$, $E_+$ the kinetic energy of the atoms or dimer after collision, and $\varepsilon$ is the kinetic energy of the released electron. If $\varepsilon_0>\text{IE}$, both PI and AI occur at any scattering energy. If both species in the initial state are  metastable states, then the energy balance Eq.~\eqref{eq:energy_balance} becomes
\begin{equation}
\label{eq:energy_balance_both_metastable}
  E_*+\varepsilon_0^A+\varepsilon_0^B=\text{IE}+\varepsilon+E_+,
\end{equation}
$\varepsilon_0^A$ and $\varepsilon_0^B$ being the excitation energy of the species $A$ and $B$, respectively. The energetic condition for the ionization then becomes $\varepsilon_0^A+\varepsilon_0^B>\text{IE}$. These conditions are fulfilled for He$^*\left(\triplets\right)$+He$^*\left(\triplets\right)$ collisions, since the first ionization energy of He is IE=24.589 and the excitation energy of $\text{He}^*(2\triplets)$ is $\varepsilon_0^A=\varepsilon_0^B=$19.820eV~\cite{NIST_ASD_2019}.

\vspace{0.6cm}
\subsection{He$^*\left(\triplets\right)$+He$^*\left(\triplets\right)$}
\label{sec:case_of_hestar_hestar_scattering}

As  described above, even at low incident kinetic energies, the electronic excitation energy of the colliding atoms allows for electron emission  in the scattering of metastable helium atoms via two different mechanisms
\begin{equation}
  \label{eq:pi_ai}
\text{He}^*+\text{He}^* \rightarrow\left\{
  \begin{array}{ll}
    \text{He}^++\text{He}+e^- & \text{(PI)} \\
    \text{He}_2^++e^- & \text{(AI)}
  \end{array}
\right.
\end{equation}
Two scenarios are possible: (i) the metastable atoms are in the same electronically excited state 2\triplets, or (ii) they are in different electronic states, in particular 2\triplets~and 2\singlets.  Here we focus on the scattering of He$^*\left(2\triplets\right)$+He$^*\left(2\triplets\right)$.  We invoke the rotating atom approximation (RAA) that assumes that the angular momentum of the colliding atoms faithfully follows the internuclear axis during the collision~\cite{Mori1964},~\ie, there is no dynamical reorientation of the {internal} angular momenta of the atoms {(here, the electronic spins of He$^*$)}
in the scattering process~\cite{Arango2006b,Arango2006a}. This approximation is very accurate in the thermal regime~\cite{Muller1991}, as demonstrated by our recent calculations of the PI and AI cross sections for Ne$^*$-Ar collisions, whose product ratio agrees with experiment over a wide range of collision energies down to 0.02~K \cite{Gordon2018}. For the case of He$^*(2^3\text{S})$+He$^*(2^3\text{S})$~\cite{Muller1991} the possible processes are

  \begin{widetext}
    \begin{subequations}
      \begin{equation}
      \label{eq:3sigmauto2sigmau}
        \text{He}^*(2\triplets)+\text{He}^*(2\triplets)\left[1^3\Sigma_u^+\right]\rightarrow \text{He}(1\singlets)+\text{He}^+(1\doublets)\left[^2\Sigma_u^+\right]+e^-,
      \end{equation}
      \begin{equation}
      \label{eq:3sigmauto2sigmag}
        \textcolor{white}{\text{He}^*(2\triplets)+\text{He}^*(2\triplets)\left[1^3\Sigma_u^+\right]}\rightarrow \text{He}(1\singlets)+\text{He}^+(1\doublets)\left[^2\Sigma_g^+\right]+e^-,
      \end{equation}
      \begin{equation}
      \label{eq:1sigmagto2sigmau}
        \text{He}^*(2\triplets)+\text{He}^*(2\triplets)\left[1^1\Sigma_g^+\right]\rightarrow \text{He}(1\singlets)+\text{He}^+(1\doublets)\left[^2\Sigma_u^+\right]+e^-,
      \end{equation}
      \begin{equation}
      \label{eq:1sigmagto2sigmag}
        \textcolor{white}{\text{He}^*(2\triplets)+\text{He}^*(2\triplets)\left[1^1\Sigma_g^+\right]}\rightarrow \text{He}(1\singlets)+\text{He}^+(1\doublets)\left[^2\Sigma_g^+\right]+e^-.
      \end{equation}
    \end{subequations}
  \end{widetext}
{The atomic} entrance channels can also couple to
 the $1^5\Sigma_g^+$ electronic state of He$_2^*$, but the autoionization of this quasimolecular state is strongly suppressed due to spin conservation. Specifically,~{since} we assume that the total spin of the collision complex is conserved, { it is not possible to obtain an exit channel with the same spin as the entrance channel $1^5\Sigma_g^+$  by coupling the states $\doubletsigmag$ or $\doubletsigmau$ with an ejected electron following the ionization.}
 Further details can be found in Refs.~\cite{Hill1972,Garrison1973,Muller1987}.  Note that the electrons in Eqs.~\eqref{eq:3sigmauto2sigmau}   and   ~\eqref{eq:1sigmagto2sigmau}, and also in Eqs.~\eqref{eq:3sigmauto2sigmag}   and  ~\eqref{eq:1sigmagto2sigmag}  are in different states, so these product channels show no interference.

Here we consider the case of  collisions of two bosonic $^4$He atoms [$\text{He}^*(2\triplets)+\text{He}^*(2\triplets)$], which  imposes certain conditions on the \wf: Namely, the orbital angular momentum $J_*$ for the collision, can  only take even (odd) for gerade (ungerade) electronic states of the collision. This effect also restricts  the angular momentum of the outgoing electron, $\ell$. Specifically, a transition from gerade (ungerade) to gerade (ungerade) leads to $\ell$ even, whereas change of gerade to ungerade or vice versa implies $\ell$ odd~\cite{Muller1991}.

\begin{figure}[h]
\centering
 \includegraphics[width=.9\linewidth]{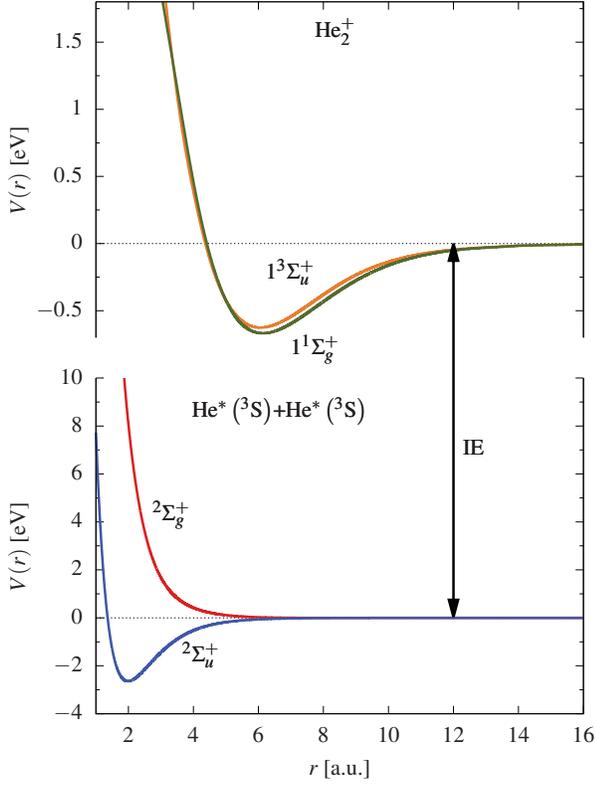}
   \caption{\label{fig:fig1} Potential energy curves $V(r)$ for  He$^*\left({}^3\text{S}\right)$-He$^*\left({}^3\text{S}\right)$ involved in the entrance and the exit channels. Note that the scale used is different for the two channels. The potentials have been taken from~\cite{Muller1991,Xie2005}, see text for further details.}
\end{figure}

To account for  autoionization, the entrance channels are described by optical potentials $V_*(R)=V(R)-\frac{i}{2}\Gamma_t(R)$~\cite{Muller1991,Garrison1973, Leo2001, Stas2006}, where $V(R)$ are the potential curves shown in Fig.~\ref{fig:fig1} and $\Gamma_t(R)$  is the total ionization width. The ionic molecular states $^2\Sigma_u^+$ and $^2\Sigma_g^+$ are described by the analytical potentials of $V_+(R)$ taken from~\cite{Xie2005}.  The real part of the optical potentials, $V(R)$, for $R\in[3,14)$ is obtained by interpolating the data in Table 3  of Ref.~\cite{Muller1991}. For $R\le 3$ we set $V(R)=V(3)$, and for $R\ge 14$  the long range expansion in terms of the $C_6$ and $C_8$ coefficients are used. The autoionization widths  {$\Gamma_t$, given by the imaginary part of $V_*(R)$}, are obtained by interpolating the data given in Ref.~\cite{Muller1991} for the interval $R\in [3,9]$. For $R\le 3$ we assume that $\Gamma(R)=\Gamma(3)$ and {we approximate the dependence $\Gamma_t(R)$ by} an exponential tail for $R\ge 9$.

 \subsection{Initial state}
\label{sec:initial_state}
The electronic state of He$^*({2}\triplets)$ is predominantly described by the configuration $1s2s$ in terms of the $1s$ and $2s$ atomic orbitals. The electronic part {of the He$^*$ \wf} is given by a linear combination of spin functions
$\ket{1M}$,
where the total spin is $1$ and $M$ is the projection of the magnetic quantum number along the $Z$ axis of the laboratory fixed frame (LFF).  Specifically,
\begin{equation}
  \label{eq:hestar_wf}
  \ket{\phi_{{2}^3S}(1,2,\left\{a\right\})}=\mathscr{A}\left\{\Phi(\vec{r}_1,\vec{r}_2)\sum\limits_{M=-1}^1a_M\ket{1M}\right\},
\end{equation}
where $\Phi(\vec{r}_1,\vec{r}_2)$ is the spatial part as a function of the position of the electrons, $a_M$ are the preparation coefficients and $\mathscr{A}$ is the antisymmetrization operator.
The initial scattering state of two He$^*$ atoms in the center-of-mass frame (CMF) is then
\begin{widetext}
  \begin{eqnarray}
    \nonumber
    \ket{\Psi\left(1,2,3,4,\{a\},\{b\},\vec{r}\right)}&=&\mathscr{A}\left\{\left[\Phi\left(\vec{r}_1-\frac{\vec{r}}{2},\vec{r}_2-\frac{\vec{r}}{2}\right)\sum\limits_{M=-1}^1a_M\ket{1M}\right]\times\right.\\
    \label{eq:hehestar_wf}
    && \left.   \left[\Phi\left(\vec{r}_3+\frac{\vec{r}}{2},\vec{r}_4+\frac{\vec{r}}{2}\right)\sum\limits_{M'=-1}^1b_{M'}\ket{1M'}\right]\right\}
  \end{eqnarray}
where $\vec{r}$ is the relative position of the He nuclei. In general, the wave function  of the  He$^*\left(2\triplets\right)$-He$^*\left(2\triplets\right)$ dimer may be written as
  \begin{eqnarray}
  \nonumber
    \ket{\Psi\left(1,2,3,4,\{a\},\{b\},\vec{r}\right)}&=&\mathscr{A}\left\{\left[\widetilde\Phi\left(\vec{r}_1,\vec{r}_2,\vec{r}_3,\vec{r}_4,\vec{r}\right)\sum\limits_{M=-1}^1\sum\limits_{M'=-1}^1a_Mb_{M'}\ket{1M}\ket{1M'}\right]\right\}=\\
    \label{eq:state_coupled}
    &=&\mathscr{A}\left\{\left[\widetilde\Phi\left(\vec{r}_1,\vec{r}_2,\vec{r}_3,\vec{r}_4,\vec{r}\right)\sum\limits_{S=0}^2\sum\limits_{M''=-2}^2c_{S,M''}\ket{SM''}_\text{He-He}\right]\right\},
  \end{eqnarray}
\end{widetext}
where~$\widetilde\Phi\left(\vec{r}_1,\vec{r}_2,\vec{r}_3,\vec{r}_4,\vec{r}\right)$ is the spatial part of the He$^*$-He$^*$ dimer electronic \wf, which depends on  the position vectors $\vec{r}_j$ of the $j$th electron and the internuclear separation $\vec{r}$ of the dimer. Also note that $\ket{SM_S}_\text{He-He}$ corresponds to the angular momentum \wf~in the molecular coupled basis.
The states with $S=0,1$ and $2$
correspond to the ${}^1\Sigma^+_g$,  ${}^3\Sigma^+_u$ and  ${}^5\Sigma^+_g$ electronic states, and the molecular coefficients $\{c\}$ are expressed via the atomic preparation coefficients as
\begin{equation}
  \label{eq:ab_to_c}
 c_{S,M''}=\sum\limits_{M=-1}^1\sum\limits_{M'=-1}^1a_Mb_{M'}\langle S M''|1M,1M'\rangle.
\end{equation}
Specifically, for the He$^*$ dimer, $S=0$, 1, and 2, and we have
\begin{subequations}
\begin{equation}
\label{eq:c22}
  c_{22}=a_1b_1,
\end{equation}
\begin{equation}
\label{eq:c21}
  c_{21}=\cfrac{1}{\sqrt{2}}(a_1b_0+a_0b_1),
\end{equation}
\begin{equation}
\label{eq:c20}
  c_{20}=\cfrac{1}{\sqrt{6}}\left(a_1b_{-1}+2a_0b_0 +a_{-1}b_1\right),
\end{equation}
\begin{equation}
\label{eq:c2m1}
  c_{2-1}=\cfrac{1}{\sqrt{2}}(a_0b_{-1}+a_{-1}b_0),
\end{equation}
\begin{equation}
\label{eq:c2m2}
  c_{2-2}=a_{-1}b_{-1},
\end{equation}
\begin{equation}
\label{eq:c11}
  c_{11}=\cfrac{1}{\sqrt{2}}(a_1b_0-a_0b_1),
\end{equation}
\begin{equation}
\label{eq:c10}
  c_{10}=\cfrac{1}{\sqrt{2}}(a_1b_{-1}-a_{-1}b_{1}),
\end{equation}
\begin{equation}
\label{eq:c1m1}
  c_{1-1}=\cfrac{1}{\sqrt{2}}(a_0b_{-1}-a_{-1}b_0),
\end{equation}
\begin{equation}
\label{eq:c00}
  c_{00}=\cfrac{1}{\sqrt{3}}(a_1b_{-1}-a_0b_0+a_{-1}b_1).
\end{equation}
\end{subequations}
The total cross section corresponding to the initial superposition of atomic states described by the \wf~$\ket{\Psi}=\sum_{S,M} c_{S,M}\ket{SM}$ at a collision energy $E_*$ may be written as
\begin{eqnarray}
\nonumber
    \sigma(\{c_{S, M}\})(E_*)&=&\sum\limits_{S, M}|c_{SM}|^2\sigma_{S,M}(E_*)+\\
    \nonumber
    &&\sum_{S\ne S'}\sum_{M\ne M'}c_{S,M}^*c_{S',M'}\sigma_{S,S',M,M'}(E_*)\\
    \label{eq:sigma_total}
    &&
\end{eqnarray}
{where $\sigma_{S,S',M,M'}(E_*)$ is the contribution to the integral cross section corresponding to the interference between the channels $S,M$ and $S',M'$~\cite{Arango2006a,Gordon2018,Shapiro2012}}
\begin{equation}
\label{eq:sigma_j_jp_mmp}
  \sigma_{S,S',M,M'}(E_*)=\SumInt\limits_q f_{S,M,q}^*(E_*)f_{S',M',q}(E_*)\mathrm{d}q.
\end{equation}
where $f_{S,M,q}(E_*)$ is the scattering amplitude for the initial state characterized by the total spin $S$ and its projection  $M$, and for the exit channel with quantum numbers $q$,~{which include  the final scattering angles of the ejected electron and of the dimer.}

Note that  $\sigma_{S,S',M,M'}(E_*) =  \sigma_{S,S',M+M'}(E_*)\delta_{M,M'}$ due to the rotational symmetry around the internuclear axis.  As discussed above,  $\sigma_{S,S',M+M'}$ are non-zero only if $S,S' \neq 2$, since the PI and AI are spin forbidden for the $^5\Sigma^+_g$ state, which corresponds to $S = 2$. Here we neglect the weak magnetic dipole-dipole interactions betwen the electronic states of different S, which is a good approximation, verified experimentally in Ref. \cite{Hill1972}.

\subsection{Considerations on internal symmetries}
\label{sec:internal_symmetries}

As  demonstrated in  previous work~\cite{Omiste2018_penning}, the symmetries of the system imply certain conditions on coherent control.
Specifically, for  He$\left(2\triplets\right)$-He$\left(2\triplets\right)$ scattering, the total final channel (molecular channel + outgoing electron) is uniquely determined by the initial scattering channel and the initial scattering energy. First, the invariance under rotations in the laboratory frame implies that two initial states interfere only if they have the same total magnetic quantum number $M$. Second, consider now the interference between pathways mediated by the parity symmetry. The initial collision state via the ionizing channels ${}^3\Sigma^+_u$ or ${}^1\Sigma^+_g$ determines that the total final channel must be \emph{ungerade} or \emph{gerade}. Therefore, even if the ionization from two different channels decays to the same molecular channel, the ejected electron carries information about the parity of the initial state, defining the total state of the system. These two conditions imply that $\sigma_{S,S',M+M'}=0$ if $S\ne S'$, where $M$ and $M'$ are the magnetic quantum number of the initial atomic states along the laboratory $Z$ axis. Thus, the cross section [Eq. {\eqref{eq:sigma_total}}] is
\begin{equation}
\label{eq:sigma_final}
  \sigma(\{c_{S,\bar M}\})(E_*)=\sum\limits_{S,\bar M}|c_{S\bar M}|^2\sigma_{S}(E_*),
\end{equation}
with $\sigma_{S}(E_*)\equiv \sigma_{S,S,\bar M}(E_*)$ { and $\bar M=M+M'$}. Note that $\sigma(\{c_{S,\bar M}\})(E_*)$ does not depend on the total magnetic quantum number {$\bar M$}, but only on the total spin $S$ of the {He$^*$-He$^*$ collision complex} and the kinetic energy.

 In other words, Eq.~\eqref{eq:sigma_final} establishes that the total cross section for a given collisional energy $E_*$ depends on the population   $|c_{S\bar M}|^2$  of each molecular channel in the initial state. We note that while Eq. (\ref{eq:sigma_final}) contains no interference terms in the molecular basis due to symmetry restrictions, there are interference  terms in the atomic basis, which is related to the molecular basis via Eq.~(\ref{eq:ab_to_c}). Thus, the cross sections (Eq. \ref{eq:sigma_final}) can be controlled by varying the relative phases of the atomic preparation coefficients $a_M$ and $b_M$ defined by Eqs.~(\ref{eq:c22})-(\ref{eq:c00})~\cite{Omiste2018_penning}.  Note that the different roles of interference contributions in the atomic and molecular basis is both insightful and unique to scattering that is reliant on vector properties.  From the viewpoint of control, however, the atomic basis, and its associated interference attributes, is the more important basis since the atomic states can be prepared in the laboratory.  

To summarize the above discussion,
the initial state of each of the colliding atoms leads to the initial state of the collision  in the form of a superposition of several molecular channels, whose population  depends only on the initial atomic states. In the case of He$^*(2\triplets)$-He$^*(2\triplets)$, the populations can be computed using the coefficients in Eqs.~\eqref{eq:c22}-\eqref{eq:c00}, which are determined by the preparation coefficients of each individual He$^*(2\triplets)$ state and, therefore, on their relative phases. In the next section we give some examples where changing the relative phase between the different initial states leads to significant changes in the cross sections, and hence control.

\section{Results}
\label{sec:results}

Here, we consider the ionization cross sections for particular entrance channels and several coherent control scenarios that can be used to benefit from  interference between scattering channels to enhance or diminish  collision cross sections.

\subsection{Absolute cross sections}
\label{sec:absolute_cross_sections}
\begin{figure}[t]
\centering
  \includegraphics[width=.9\linewidth]{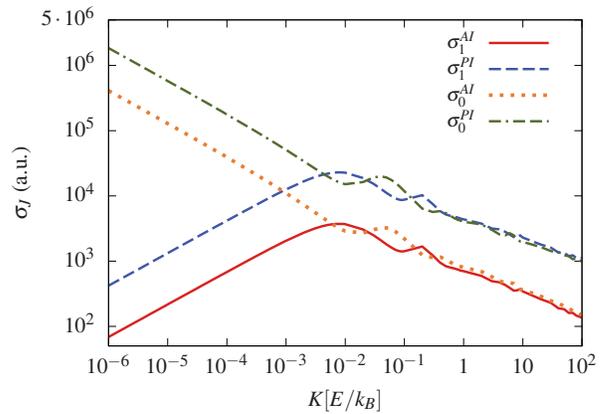}
  \caption{\label{fig:fig2} AI and PI cross sections for the $S=0$ and $1$ channels as a function of  temperature for He$^*$(2\triplets)-He$^*$(2\triplets) scattering. Recall that $\sigma_0$ applies to the $\singletsigma^+$ channel and $\sigma_1$ to the $\tripletsigma^+$.}
\end{figure}

We first compute the ionization cross sections $\sigma_S$ of Eq.~(\ref{eq:sigma_final}) for the processes given by Eqs.~\eqref{eq:3sigmauto2sigmau}-\eqref{eq:1sigmagto2sigmag} in the molecular channels $\singletsigma^+$ and $\tripletsigma^+$. The total $S$-dependent AI and PI cross sections $\sigma_0$ and $\sigma_1$, are shown in Fig.~\ref{fig:fig2}. The PI and AI for both of the autoionizing channels have similar values at temperatures higher than 10~mK, although for the same type of ionization (\ie, AI or PI) there are \emph{out-of-phase} oscillations for $\singletsigma^+$ and $\tripletsigma^+$. This pattern appears because both entrance molecular channels are similar and the high number of partial waves that contribute allows access to He($\singlets$)-He$^+$($\singlets$). {Note that the} cross sections in this regime can be improved by using a more sophisticated $V_{\epsilon\ell}(r)$ for non s-wave ejected electrons{, which accounts for steric effects in autoionization}.

However, for temperatures below 10~mK,  autoionization presents a very different pattern for the $\singletsigma^+$ and $\tripletsigma^+$ channels, as shown in Fig.~\ref{fig:fig2}. On the one hand, the autoionization for the entrance channel $\singletsigma^+$ becomes much larger at low temperatures~{in accord with the Wigner threshold law}. In this regime, only the partial waves with low incident angular momenta contribute,   which are $J_*=J_+=0$ for Eq. ~\eqref{eq:1sigmagto2sigmag} and $J_*=0$ and $J_+=1$ for~Eq. \eqref{eq:1sigmagto2sigmau}, since the $J_+=0$ is forbidden for gerade channels in the bosonic case, as described in Sec.~\ref{sec:internal_symmetries}. The overlap between the initial $\singletsigma^+$ scattering state  and the outgoing scattering states of $\doubletsigmag^+$ symmetry is larger than between the states $\singletsigma^+$ and $\doubletsigmau^+$. This  can be explained for the outgoing continuum states in terms of the position of the short-range barrier, which is closer to the nucleus for $\doubletsigmau^+$ than for $\doubletsigmag^+$. This creates a larger  overlap with the entrance scattering states  for the $\doubletsigmag^+$ state. It is counterintuitive that the AI is larger for the gerade exit channel whereas we find only 2 bound states for $J_+=0$ vs. 26 for $\doubletsigmau^+$. This can be explained by noting that the bound states of the $\doubletsigmau^+$ channel are  located at $R\simeq 3$~a.u., far from  the entrance channel minimum, which occurs at $R\simeq6$~a.u., leading to a small overlap. On the contrary, the bound states supported by the $\doubletsigmag^+$ electronic state are located around 8~a.u., having a significant overlap.

{On the other hand}, the ionization through the $\tripletsigma^+$ state occurs mainly due to $J_*=1$ partial wave in the ultracold regime, which implies that the atoms encounter a centrifugal barrier. Therefore, the lower  the temperature, the lower  the probability to tunnel through the barrier. This effect diminishes the overlap, and therefore the ionization cross sections decrease, as shown in Fig.~\ref{fig:fig2}. These considerations are not relevant for higher temperatures {because the real part of the ionic potentials, shown in Fig.~\ref{fig:fig1},  differ only slightly around the well. As a result, the scattering states are very similar for all the exit channels. Besides, this well is located close to a region where the entrance potentials are very similar, which implies a similar overlap~{of the incoming scattering~\wf~of the $\tripletsigma^+$ and $\singletsigma^+$ states with the bound and scattering states of $\doubletsigmag$ and $\doubletsigmau$ symmetries}
in that region. As mentioned above, we observe an~\emph{out of phase} oscillation  in the cross section in Fig.~\ref{fig:fig2} above $10$~mK, induced by the phase shift determined by the inner part of the potential.} We note that the autoionization rates from the entrance channels to the two accessible exit channels are very similar for the optical potentials and $V_{\epsilon\ell}(R)$ considered.

\subsection{Rotated states}
\label{sec:rotated_states}
First, we consider the rotation of states with well defined magnetic quantum numbers as a theoretical  protocol to prepare the initial atomic states. The selection of atomic states  $|1M\rangle$ with well-defined quantum numbers as well as their rotation by means of external magnetic fields allows for an efficient preparation of initial states for coherent control with currently available technologies
\cite{Gordon2018,Zou2018}. The rotation operator $R_\theta^Y(\theta)$ applied to the initial two-atom  state $\ket{1M_A}\ket{1M_B}$ gives
\begin{equation}
  \label{eq:rotate_jm}
  R_\theta^Y\ket{1M}=\sum\limits_{M'=-1}^1d_{M',M}^1(\theta)\ket{1M'},
\end{equation}
where $R_\theta^Y$ denotes a rotation by angle $\theta$ around the $Y$-axis of the laboratory fixed frame and $d_{M',M}^S(\theta)$ are the reduced Wigner matrix elements~\cite{Zare1988}. Consider then the initial state $\ket{1M_A}\ket{1M_B}$ and consider the cross sections after rotating atom $A$ by an angle $\alpha$ and  atom $B$ an angle $\beta$. The initial rotated state of the two atoms then becomes, as a function of the control parameters $\alpha$ and $\beta$
\begin{widetext}
  \begin{equation}
    \label{eq:rotate_jma_jmb}
    R_\alpha^Y\ket{1M_A}R_\beta^Y\ket{1M_B}=\sum\limits_{M_A'=-1}^1\sum\limits_{M_B'=-1}^1d_{M_A',M_A}^1(\alpha)d_{M_B',M_B}^1(\beta)\ket{1M_A'}\ket{1M_B'}.
  \end{equation}
\end{widetext}
We compute the PI and AI cross sections for this superposition of initial states using Eq. (\ref{eq:sigma_final}) in combination with Eq.~\eqref{eq:ab_to_c}, which allows us to compute the preparation coefficients in the molecular basis.

\begin{figure}
    \centering
    \includegraphics{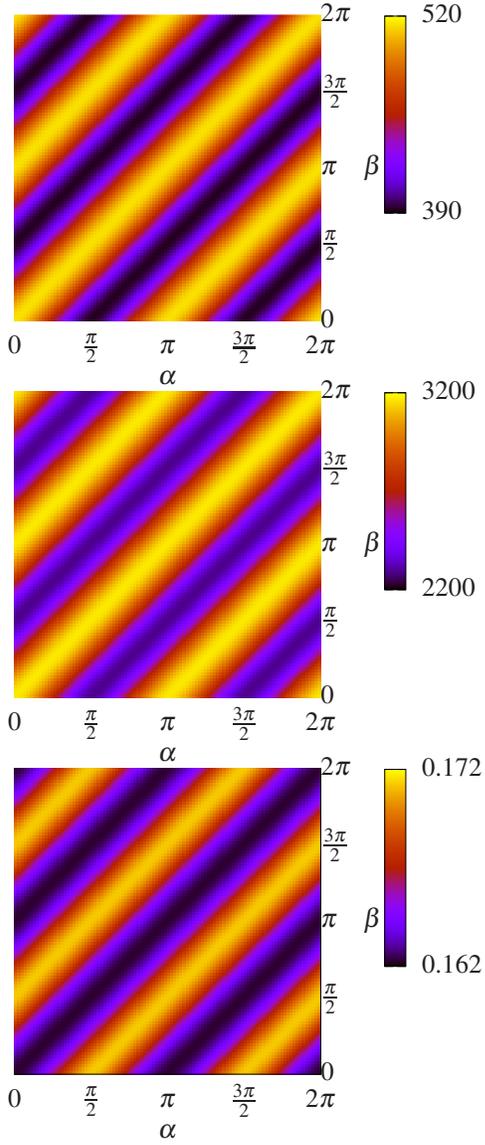}
    \caption{\label{fig:fig3} Scattering results for the entrance channel $R_Y(\alpha)R_Y(\beta)\ket{11}\ket{10}$ at 10~$mK$; the $\sigma^{AI}$ (upper panel), $\sigma^{PI}$ (middle panel) and ${\sigma^{AI}}/{\sigma^{PI}}$ (lower panel).}
\end{figure}

\begin{figure}[b]
    \centering
    \includegraphics{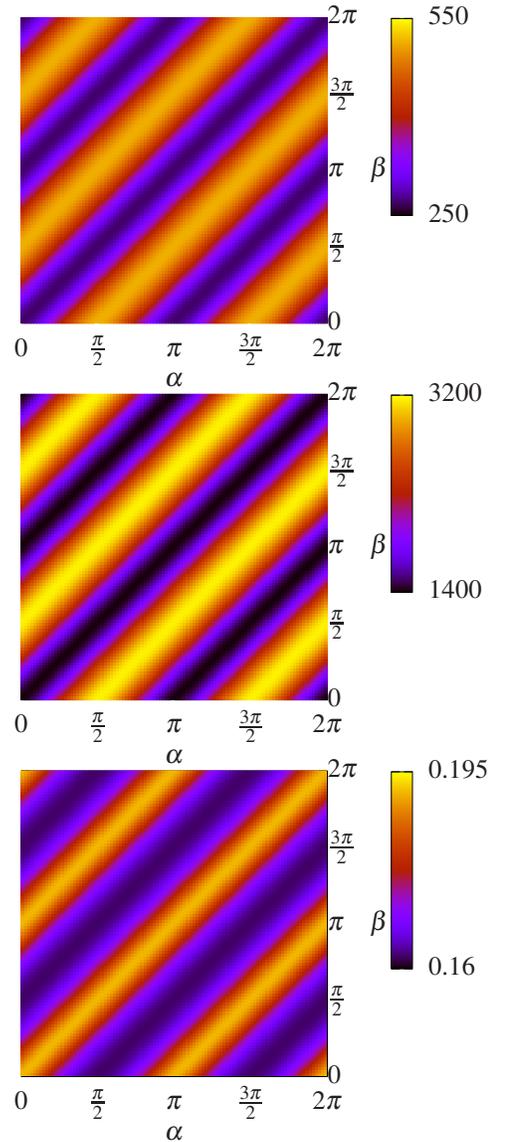}
    \caption{\label{fig:fig4} Scattering results for the entrance channel $R_Y(\alpha)R_Y(\beta)\ket{10}\ket{10}$ at 10~$mK$; the $\sigma^{AI}$ (upper panel), $\sigma^{PI}$ (middle panel) and ${\sigma^{AI}}/{\sigma^{PI}}$ (lower panel).}
\end{figure}

 Figures~\ref{fig:fig3} and~\ref{fig:fig4} show the ionization cross sections $\sigma^{AI}$, $\sigma^{PI}$ and their ratio as a function of the control angles $\alpha$ and $\beta$ for the initial two-atom states $\ket{11}\ket{11}$, $\ket{11}\ket{10}$  and $\ket{10}\ket{10}$ rotated according to Eq. (\ref{eq:rotate_jma_jmb}),  at 10~mK.
  The cross sections are characterized by  a  pronounced band structure, which manifests itself in the dependence on the relative atomic orientation, $\gamma=\alpha-\beta$.


This band structure is due to the independence of the cross section $\sigma_S$ on the total magnetic quantum number $M$, which  only determines the population of the $\singletsigma^+,~\tripletsigma^+$ and $\quintetsigma^+$ states via the preparation coefficients $c_{SM}$ in Eq.~\eqref{eq:sigma_final}. We can extend this argument to any rotation around the $Y$-axis applied to both atoms simultaneously. However, if  the Hamiltonian includes terms that couple the translational and rotational degrees of freedom, such as the spin-spin coupling, this phenomenon does not hold, as in the case of Ne$^*$-Ar scattering~\cite{Omiste2018_penning,Gordon2018}. Further details can be found in Appendix~\ref{sec:relative_orientation}.

\begin{figure*}
\centering
  \includegraphics[width=\linewidth]{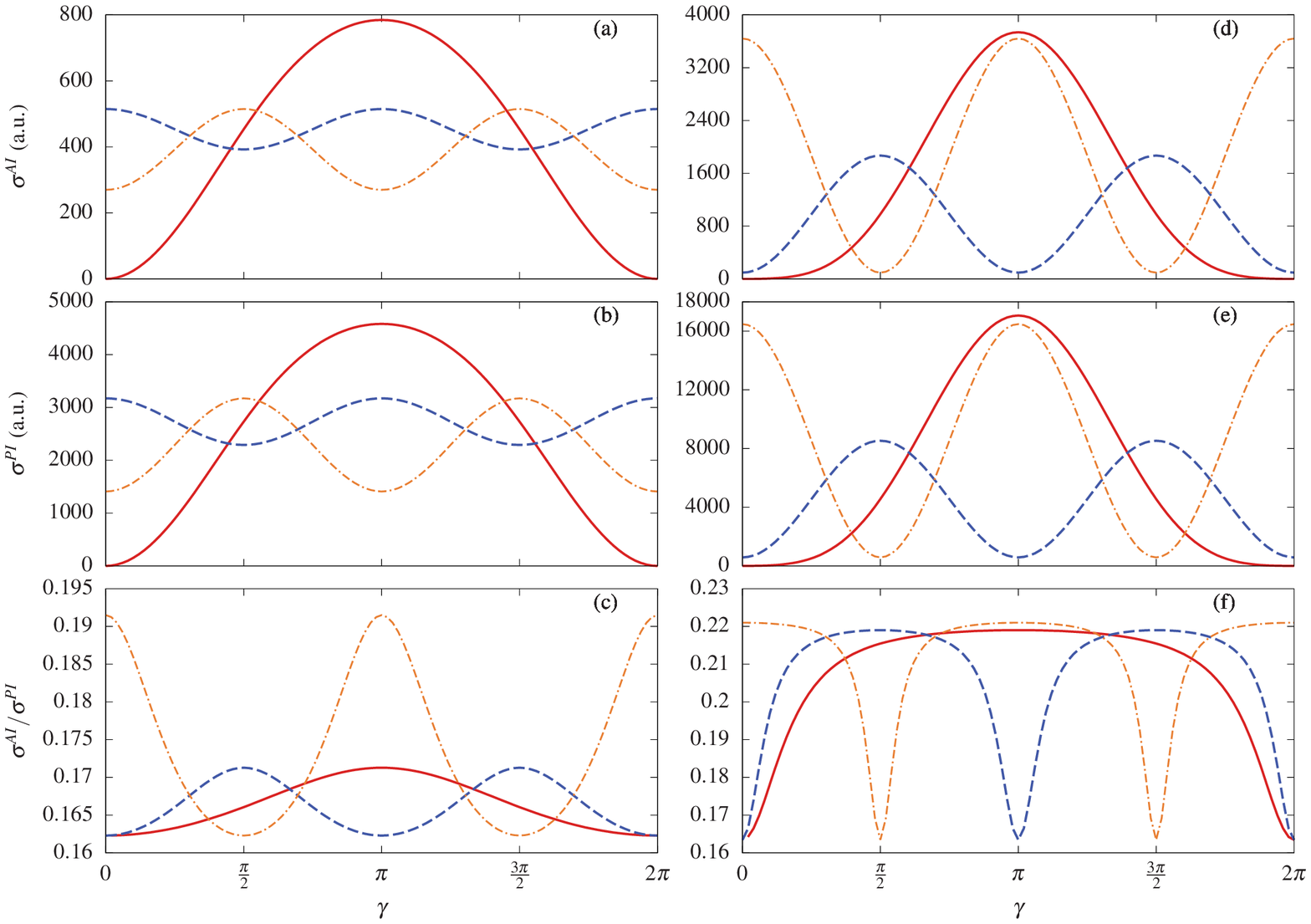}
  \caption{\label{fig:fig5} Cross sections at  $T=10$~mK (a) $\sigma^\text{AI}$, (b) $\sigma^\text{PI}$ and (c) ionization ratio $\sigma^\text{AI}/\sigma^\text{PI}$ and cross sections at  $T=100$~$\mu$ K (d) $\sigma^\text{AI}$, (e) $\sigma^\text{PI}$ and (f) ionization ratio $\sigma^\text{AI}/\sigma^\text{PI}$. Results are for rotations around the $Y$~axis as a function of $\gamma=\beta-\alpha$ for the states $\ket{11}\ket{11}$ (solid red), $\ket{11}\ket{10}$ (dashed blue) and $\ket{10}\ket{10}$ (dash-dotted orange). Note that we assume $\ell_\text{max}=1$.}
\end{figure*}

In Figure~\ref{fig:fig5} we  show the absolute cross sections and the ionization ratio for the initial scattering states $\ket{11}\ket{11}$, $\ket{11}\ket{10}$ and $\ket{10}\ket{10}$ at $10~mK$ as a function of $\gamma=\alpha-\beta$ at 100~$\mu$K. It is noteworthy that the patterns of the ionization ratio for $\ket{10}\ket{10}$ and $\ket{11}\ket{10}$ are different as the collision energy is lowered from 10~mK to 100~$\mu$K. As an example, consider AI for $\ket{11}\ket{10}$ shown in Fig.~\ref{fig:fig5}(a) and Fig.~\ref{fig:fig5}(d). For 10~mK we observe a maximum of $\sim 520$ at $\alpha=\beta=0$, which diminishes to $\sim$390 at $\alpha=0$~and $\beta=\pi/2$. On the other hand, this pattern is inverted for $100~\mu K$, where $\sigma^{AI}$ has a minimum at $\alpha=\beta=0$ and a maximum that is reached by increasing $\beta$ to $\pi/2$. By comparing the cross section at these two configurations we find that the maximum is located at $\alpha=\beta=0$ if $3\sigma_1> 2\sigma_0$, as is the case for 10~mK. However, we observe that the maxima and the minima are located in the same regions of $\gamma$, although they can interchange their role. In the case of $\ket{11}\ket{11}$, the pattern is found to be independent of the scattering energy, since for $\alpha=\beta=0$ the cross sections for both AI and PI are zero. The location and characteristics of the maxima and minima are discussed in detail  in Sec.~\ref{sec:designed_states}.

{Figures~\ref{fig:fig5}(c) and Figures~\ref{fig:fig5}(e)}
show the ratio $\sigma^\text{AI}/\sigma^\text{PI}$, which is dramatically different at 10~$mK$ and 100~$\mu K$.
The control of AI and PI as a function of angle $\gamma=\beta-\alpha$, shown in Fig.~\ref{fig:fig5}, is significant. Of particular relevance is the ratio of the two product channels since the output of either product channel could be varied by varying the incident flux
For example, the $\ket{11}\ket{10}$ case at 10~$mK$ [Fig.~\ref{fig:fig5}(a)-(c)] shows a variation of the product ratio $\sigma^\text{AI}/\sigma^\text{PI}$ from $0.19$ to $0.162$,~\ie a $19\%$ variation. At 100~$\mu K$ this range increases from $0.165$ to $0.22$, a $33\%$ variation.

{Finally, note that the ratio $\sigma^\text{AI}/\sigma^\text{PI}$ is not defined for the initial state $\ket{11}\ket{11}$, since it only couples to the $\quintetsigma^+$ channel, which does not autoionize. However, we can identify this ionization ratio with $\sigma^\text{AI}_1/\sigma^\text{PI}_1$ corresponding to the dominant contribution when rotating a small angle around the $Y$-axis, which also matches the ratio for the initial state $\ket{11}\ket{00}$, as illustrated in Fig.~\ref{fig:fig5}.}

\subsection{Searching for maximum cross sections}
\label{sec:designed_states}
In this section, we explore the extent of the coherent control of PI and AI in cold He$^*$-He$^*$ collisions.  To this end, we search for the initial  parameters $\alpha$ and $\beta$, that maximize (minimize) the PI and AI cross sections as well as the ionization ratio $\sigma^{AI}/\sigma^{PI}$.
As noted above, these states can be prepared experimentally by  standard techniques such as coherent population transfer (CPT)~\cite{Arimondo1996}, electromagnetically induced transparency (EIT)~\cite{Fleischhauer2005} or stimulated Raman adiabatic passage (STIRAP)~\cite{Vitanov2017}.

Consider first the maxima and minima of the individual AI and PI cross sections, $\sigma$ [Eq.~\eqref{eq:sigma_final}]; details of the optimization method are provided in Appendix~\ref{sec:optimization_sigma}.
We obtain that the preparation coefficients fulfill the conditions 
\begin{subequations}
\begin{equation}
\label{eq:self_consistent_a}
    a_j=\dfrac{\sum_{SM}\sigma_{SM} c_{SM}\dfrac{\partial c_{SM}^\star}{\partial a_j^\star}}{\sum_{SM}|c_{SM}|^2\sigma_{SM}}.
\end{equation}
\begin{equation}
\label{eq:self_consistent_b}
    b_j=\dfrac{\sum_{SM}\sigma_{SM} c_{SM}\dfrac{\partial c_{SM}^\star}{\partial b_j^\star}}{\sum_{SM}|c_{SM}|^2\sigma_{SM}}.
\end{equation}
\end{subequations}
We can check by inspection that the following set of four preparation coefficients
\begin{widetext}
    \begin{eqnarray}
    \label{eq:solution1}
     a_0=b_{0}=1, a_1=a_{-1}=b_1=b_{-1}=0&\Rightarrow& \sigma=\dfrac{\sigma_0}{3},\\
        \label{eq:solution2}
 a_0=b_{-1}=1, a_1=a_{-1}=b_1=b_0=0&\Rightarrow&
 \sigma=\dfrac{\sigma_1}{2},\\
     \label{eq:solution3}
    a_1=b_{-1}=1, a_0=a_{-1}=b_1=b_0=0&\Rightarrow& \sigma=\dfrac{\sigma_1}{2}+\dfrac{\sigma_0}{3},\\
         \label{eq:solution4}
     a_1=b_{1}=1, a_0=a_{-1}=b_{-1}=b_0=0&\Rightarrow& \sigma=0.
    \end{eqnarray}
    \end{widetext}
are solutions of Eqs.~\eqref{eq:self_consistent_a} and~\eqref{eq:self_consistent_b}, and thus, maximize or minimize  the total cross sections:  the first two provide local extrema whereas the second two are absolute maxima and minima, respectively.  

Note that $\sigma$ in Eqs.~\eqref{eq:solution1}-\eqref{eq:solution4}  apply to both PI and AI.  For example, Eq.~\eqref{eq:solution1} indicates that $\sigma^{AI} = \sigma^{AI}_{0}/3$ and  $\sigma^{PI} = \sigma^{PI}_{0}/3$.  Further, note that the minimum of zero in Eq.~\eqref{eq:solution4} results from the fact that all population is in $S = 2$ where autoionization is forbidden.

We also find that the preparation coefficients~\eqref{eq:solution3} fulfill the following more general conditions
    \begin{eqnarray}
    \nonumber
            &&|a_1+a_{-1}|^2=|b_1+b_{-1}|^2=1,\\
            \label{eq:solution5}
            &&a_j=e^{i\rho}(-1)^{j}b_{-j},\quad\rho\in [0,2\pi)\\
            \nonumber
            &&\dfrac{a_j}{a_k}, \dfrac{b_j}{b_k}\in\mathds{R}\quad\text{for }j, k=-1,0,1,
    \end{eqnarray}

    Note that the  optimal preparation coefficients correspond to eigenstates of the symmetry operator that describe the collision,~\ie, the arbitrary rotations around the internuclear axis. Therefore, from the coefficients defined by Eqs.~\eqref{eq:solution1}-\eqref{eq:solution5}  we can generate infinitely many solutions by means of arbitrary rotations around the internuclear axis, $\vec{k}$, two-fold rotations perpendicular to $\vec{k}$ and interchange of the nuclei.       In addition, we also find that performing the same rotation in both atoms does not change the cross section, because the potentials depend on $S$ of the molecular channels, but not on the value of $M$, as described in Sec.~\ref{sec:rotated_states} (see also Appendix~\ref{sec:relative_orientation}). This last transformation manifests itself as  an accidental degeneracy of the cross section, which is not related to a symmetry operation. Therefore, it can be used to check the validity of our model and to experimentally quantify the effect on control of the various  interactions neglected here, such as the spin-orbit coupling and the magnetic dipole-dipole interaction.

  As shown in Appendix \ref{sec:optimization_sigma}  the maxima and minima of the ratio ${\sigma^{AI}}/{\sigma^{PI}}$ {are obtained with the same preparation coefficients that optimize the absolute cross sections.  The coefficients (20)   and their transformations give an ionization ratio

\begin{equation}
    \label{eq:ratioj0j1}
    \dfrac{\sigma^{AI}}{\sigma^{PI}}=\dfrac{2\sigma^{AI}_0+3\sigma^{AI}_1}{2\sigma^{PI}_0+3\sigma^{PI}_1}.
\end{equation}
As in the cases considered previously, the set of preparation coefficients, which results in an optimal cross section (either maxima or minima) is independent of the collision energy. This is because the population of the autoionizing molecular channels  depends only on the internal states of each colliding helium atom since the dynamical reorientation of the atomic spin of the individual atoms is not allowed in the RAA. On the other hand, an optimal point would be either a maximum or a minimum,  depending on the cross section of each molecular channel.
In addition the preparation coefficients that optimize the
absolute cross sections and the ionization ratio are 
independent of $\sigma_0$ and $\sigma_1$, and therefore 
the optimal preparation coefficients are the same for 
Associative and Penning ionization.

Straightforward arithmetic shows that 
the right hand side of Eq. \ref{eq:ratioj0j1} implies that 
${\sigma^{AI}}/{\sigma^{PI}}$ lies between 
$\sigma_1^{AI}/\sigma_1^{PI}$ and $\sigma_0^{AI}/\sigma_0^{PI}$ at
all temperatures. 
Results for the maximum and minimum values of
$\sigma_1^{AI}/\sigma_1^{PI}$ and $\sigma_0^{AI}/\sigma_0^{PI}$ that
bound ${\sigma^{AI}}/{\sigma^{PI}}$ are shown in Fig.
\ref{fig:fig6}.

\begin{figure}[t]
\centering
  \includegraphics[width=.9\linewidth]{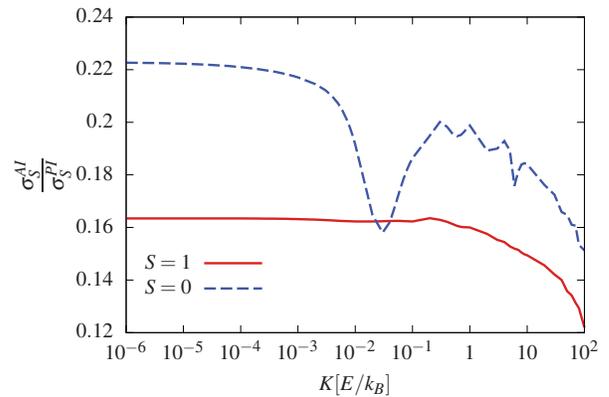}
  \caption{\label{fig:fig6} For the He$^*$(2\triplets)-He$^*$(2\triplets) scattering, absolute maximum  and minimum of the ionization ratio as a function of the temperature. In accord with the discussion in the text, the controlled cross section at any temperature lies between the maximum and minimum values shown here. }
\end{figure}

\section{Conclusions and outlook}
\label{sec:conclusions_and_outlook}
In this work, taking the collision of two metastable ${}^4$He$^*(\triplets)$ atom as a prototypical example, we discussed coherent control of ultracold scattering involving two open-shell metastable atoms. First, it is shown that the interference between the channels involved in the collision can enhance or suppress the Penning or associative ionization{, which is especially remarkable in the cold regime where the latter is more likely.} The interference can be manipulated by modifying the phase between internal states of each individual atom, allowing us to control the absolute cross section and ionization ratios. {For instance, in the ${}^4$He$^*(\triplets)$-${}^4$He$^*(\triplets)$ scattering case, we find that by tuning the preparation coefficients we are able to diminish the ionization ratio from 0.22 to 0.16}. Furthermore, we propose a computational method to  maximize and minimize the cross sections and prove that a given solution can be generalized by rotating the reference frame of the system. In particular, we find that there is one absolute maximum and one absolute minimum value for the cross sections, and that the corresponding preparation coefficients are independent of the scattering energy.

This study can be useful for future experiments on coherent control of scattering processes in the ultracold regime and can be be extended to other systems such as Rb-He$^*$ or open-shell molecules colliding with He* atoms. From the theoretical point of view, it is important to compute the optical and real potentials in the region close to the nuclei in order to accurately  describe  the reactions in the ultracold regime and possible non-isotropic contributions.

\begin{acknowledgments}
This work was supported by U.S. Air Force Office of Scientific Research under grant number FA9550-19-1-0312. J.J.O also acknowledges the funding from Juan de la Cierva - Incorporaci\'on program granted by Ministerio de Ciencia e Innovaci\'on (Spain). T.V.T. was partially supported by the NSF (Grant No.~PHY-1912668).
\end{acknowledgments}

\appendix
\section{Scattering states and cross sections}
\label{sec:scattering_states_xs}

Here, we briefly describe the methodology to compute the total PI and AI cross sections from the scattering states. We follow the same procedure as in Ref.~\cite{Arango2006a,Arango2006b,Gordon2018, Omiste2018_penning}. Work is carried out  in the center of mass frame (CMF), where the total linear momentum $\vec{K}=0$. This assumption can be taken without loss of generality, because the total cross section is the same at all inertia frames~\cite{Taylor1972, Arango2006a, Arango2006b, Omiste2018_penning}. The initial relative momentum in the internuclear axis, $\vec{k}$, is set to define the $Z$-axis of the lab, $k_f$ and $k_\varepsilon$ are the momentum of the atoms after the collision and the momentum of the Penning electron with energy $\varepsilon$, respectively.
The scattering amplitude in the Born-Oppenheimer approximation is given by~\cite{Arango2006b}
\begin{equation}
  f\left(\hat{{k}}_f,\hat{{k}}_\varepsilon;\vec{k}\right)=-\cfrac{2\mu \rho_\varepsilon^{1/2}}{\left(4\pi\hbar^2\right)}\left(\cfrac{k_f}{k}\right)^{1/2}\melement{\psi_{v,\varepsilon}}{V_{\varepsilon,\hat{{k}}_\varepsilon}}{\psi_d},
\end{equation}
where $\mu=m_\text{He}/2$ is the reduced mass and $\rho_\varepsilon$ is the density of electronic continuous states. $\psi_d(\vec{R})$ is the incoming \wf, and can be expanded as
\begin{widetext}
\begin{equation}
  \psi_d(\vec{R})=4\pi\sum_{J_*=0}^\infty\sum_{m=-J_*}^{J_*} i^le^{i\delta^{J_*}}\cfrac{\psi^{J_*}_d(R)}{k^{1/2}R}Y_{J_*m}^*(\hat{{k}})Y_{J_*m}(\hat{R})
\end{equation}
where $\delta^{J_*}$ is the complex phase and $\psi_d^{J_*}(R)$ satisfies the  Schr\"odinger equation
\begin{equation}
\left(-\cfrac{\hbar^2}{2\mu}\cfrac{\mathrm{d}^2}{\mathrm{d}R^2}+V_*(R)-\cfrac{i}{2}\Gamma_t(R)+\cfrac{\hbar^2}{2\mu}\cfrac{J_*(J_*+1)}{R^2}-E_*\right)\psi_{d,E_*}^l(R)=0,
  \label{eq:schrodinger_entrance}
\end{equation}
and has the asymptotic behavior
\begin{equation}
  \label{eq:psid_asymptotic}
  \psi_d^{J_*}(R)\xrightarrow{R\rightarrow \infty} k^{-1/2}\sin(kR-\pi J_*/2+\delta^{J_*})
\end{equation}
The scattering state for the exit channel
 $\psi_{\epsilon}(\vec{R})$ can be written as
\begin{equation}
  \psi_\epsilon(\vec{R})=4\pi\sum_{J_+=0}^\infty\sum_{m'=-J_+}^{J_+}i^{J_+}e^{-i\delta^{J_+}_f}\cfrac{\psi^{J_+}_\epsilon(R)}{k_f^{1/2}R}Y_{J_+m'}^*(\hat{{k}})Y_{J_+m'}(\hat{R}),
\end{equation}
where $\psi^{J_+}_\epsilon(R)$ fulfils the equation
\end{widetext}
\begin{equation}
  \left(-\cfrac{\hbar^2}{2\mu}\cfrac{\mathrm{d}^2}{\mathrm{d}R^2}+V_+(R)+\cfrac{\hbar^2}{2\mu}\cfrac{J_+(J_++1)}{R^2}-\epsilon\right)\psi_{v/\epsilon}^{J_+}(R)=0,
  \label{eq:schrodinger_final}
\end{equation}
where we also include the bound states $\psi_v^{J_+}(R)$. The asymptotic behavior of $\psi_\epsilon^{J_+}(R)$
\begin{equation}
  \label{eq:psie_asymptotic}
  \psi_\epsilon^{J_+}(R)\xrightarrow{R\rightarrow \infty} k_f^{-1/2}\sin(k_fR-\pi {J_+}/2+\delta_f^{J_+}),
\end{equation}
where $\delta_f^{J_+}$ is the real phase shift.
Then, assuming the $\vec{k}$ is parallel to the $z$-axis of the CMF, the scattering amplitudes for a given entrance and exit channel can be rewritten as~\cite{Arango2006a, Arango2006b}
\begin{widetext}
\begin{eqnarray}
\nonumber
f(k,k_f,k_\varepsilon)&=& \cfrac{\sqrt{\pi}}{ik}\sum_{\ell,m,J_*,J_+}i^{J_*-J_+}(2J_*+1)(2J_++1)^{1/2}\left(
  \begin{array}{ccc}
    J_+ & \ell & J_*\\
    0  & 0 & 0\\
  \end{array}
  \right)\left(
  \begin{array}{ccc}
    J_+ & \ell & J_*\\
    -m & m & 0\\
  \end{array}
  \right)
S_{J_+\ell}^{J_*}(\epsilon)Y_{J_+-m}(\hat{{k}_f})Y_{\ell m}(\hat{{k}_\varepsilon})\\
&&
\end{eqnarray}
\end{widetext}
where we include the angular momentum ($\ell$) and magnetic quantum number ($m$) of the ejected electron. The $S$ matrix is given in terms of the phase shifts $\delta^{J_*}$ and  $\delta^{J_+}_f$~\cite{Arango2006a,Arango2006b}
\begin{equation}
  S_{J_+,\ell}^{J_*}(\epsilon)=-2i\cfrac{2\mu\rho_\varepsilon^{1/2}}{\hbar^2}e^{i\left(\delta^{J_*}+\delta_f^{J_+}\right)}\melement{\psi_{\epsilon}^{J_+}}{V_{\varepsilon\ell}}{\psi_d^{J_*}},
\end{equation}
for the PI and
\begin{equation}
  S_{J_+,\ell}^{J_*}(\epsilon)=-2i\left(\cfrac{2\mu\rho_\epsilon}{\hbar^2}\right)^{1/2}e^{i\delta^J_*}\melement{\psi_{v}^{J_+}}{V_{\epsilon\ell}}{\psi^{J_*}_d},
\end{equation}
for the AI, where $V_{\epsilon\ell}(R)\approx\alpha_{\ell_{\max}}\sqrt{\Gamma_t(R)/(2\pi)}$, where $\Gamma_t(R)$ is the autoionization width and $\alpha_{\ell_{\max}}=1/\sqrt{\ell_{\max}+1}$~\cite{Khan1991}.

\section{Cross section depends on the relative orientation}
\label{sec:relative_orientation}

Here, we show  that the angular dependence of the cross section in ${}^4$He$(\triplets)$-${}^4$He$(\triplets)$ scattering only depends on the relative orientation of their atomic states.

First, we prove that a global rotation of the molecular channel does not affect the cross section. Let us define the state in the molecular entrance channel basis
\begin{equation}
    \label{eq:state_molecular_basis}
    \ket{\Psi}=\sum_{SM}c_{SM}\ket{SM},
\end{equation}
where $\ket{SM}$ denote the molecular entrance channels with $J=0, 1$ and $2$ ($\singletsigma^+,~\tripletsigma^+~\text{and}~\quintetsigma^+$, respectively) and $M$ is the total magnetic quantum number. An arbitrary rotation of $\ket{\Psi}$  is given by~\cite{Zare1988}
\begin{eqnarray}
    \nonumber
    \ket{\Psi}_{(\phi,\theta)}=R(\phi,\theta)\ket{\Psi}&=&\sum_{SM}\sum_{M'}D^S_{M',M}(\phi,\theta,0)c_{SM}\ket{SM'},\\
    \label{eq:state_molecular_basis_rotated}
    &&
\end{eqnarray}
where $(\phi,\theta)$ are the Euler angles which define the rotation, and the expansion coefficients after the rotation are given by
\begin{equation}
\label{eq:cjm_rotated}
    c_{SM'}(\phi,\theta)=\sum_{M}D^S_{M',M}(\phi,\theta,0)c_{SM} .
\end{equation}
From Eq.~\eqref{eq:sigma_final} we know that the cross section depends on the population of each molecular entrance channel. Substituting  the expansion coefficients of $\ket{\Psi}_{(\phi,\theta)}$ into Eq.~\eqref{eq:sigma_final}, we obtain
\begin{eqnarray}
    \nonumber
    &&\sigma(\{c_{SM}(\phi,\theta)\})(E_*)= \sum_{SM}|c_{SM}(\phi,\theta)|^2\sigma_S(E_*)= \\
    \nonumber
    && \sum_{SM}\sum_{M',M''}D_{M,M'}^S(\phi,\theta)D_{M,M''}^S(\phi,\theta)^\star c_{SM'}c_{SM''}^\star\sigma_S(E_*)=\\
    \label{eq:sigma_rotated}
    &&  \sum_{SM}|c_{SM}|^2\sigma_S(E_*)=\sigma(\{c_{SM}\})(E_*),
\end{eqnarray}
where we have used $\sum_M D_{M,M'}^S(\phi,\theta)D_{M,M''}^S(\phi,\theta)^\star=\delta_{M',M''}$~\cite{Zare1988}.

Applying the same rotation to the electronic state of both He$^*$ atoms is equivalent to performing a rotation to the internal state of the whole molecular system. Rotating the He atoms the same angle does not change the cross section if it  depends only on the total angular momentum of initial molecular state, i.e., the cross section is determined by the relative orientation of the atoms and the scattering energy.

\section{Optimization of the cross sections}
\label{sec:optimization_sigma}
In this section we derive the conditions that the preparation coefficients fulfill to maximize (minimize) the cross sections and the ionization ratio.

Using Lagrange's multipliers we find that the maxima and minima fulfill
\begin{widetext}
\begin{subequations}
\begin{equation}
\label{eq:critical1}
  \frac{\partial}{\partial a_j}\left[\sum_{SM}|c_{SM}|^2\sigma_{SM}-\lambda (|a_{-1}|^2+|a_{0}|^2+|a_{1}|^2-1) \right]=0,\qquad j=-1,0,1,
  \end{equation}
  \begin{equation}
  \label{eq:critical2}
  \frac{\partial}{\partial b_j}\left[\sum_{SM}|c_{SM}|^2\sigma_{SM}-\gamma (|b_{-1}|^2+|b_{0}|^2+|b_{1}|^2-1) \right]=0,\qquad j=-1,0,1,
  \end{equation}
  \begin{equation}
  \label{eq:critical3}
|a_{-1}|^2+|a_{0}|^2+|a_{1}|^2=1,
\end{equation}
\begin{equation}
\label{eq:critical4}
|b_{-1}|^2+|b_{0}|^2+|b_{1}|^2=1.
\end{equation}
\end{subequations}
\end{widetext}
 By taking derivatives $a_j~(b_j)$ and multiplying by $a_j~(b_j)$ it is easy to show that
 \begin{equation}
     \lambda=\gamma=\sum_{SM}|c_{SM}|^2\sigma_{J}=\sigma(\{c_{SM}\}).
 \end{equation}
   Therefore, {we obtain}
\begin{subequations}
\begin{equation}
\label{eq:self_consistent_a_appendix}
    a_j=\dfrac{\sum_{SM}\sigma_{SM} c_{SM}\dfrac{\partial c_{SM}^\star}{\partial a_j^\star}}{\sum_{SM}|c_{SM}|^2\sigma_{SM}}.
\end{equation}
\begin{equation}
\label{eq:self_consistent_b_appendix}
    b_j=\dfrac{\sum_{SM}\sigma_{SM} c_{SM}\dfrac{\partial c_{SM}^\star}{\partial b_j^\star}}{\sum_{SM}|c_{SM}|^2\sigma_{SM}}.
\end{equation}
\end{subequations}

On the other hand, to compute the maxima (minima) of the ionization ratio $\sigma^{AI}/\sigma^{PI}$.
    \begin{widetext}
\begin{subequations}
\begin{equation}
\label{eq:optimize_ratio1}
  \frac{\partial}{\partial a_j}\left[\dfrac{\sum_{SM}|c_{SM}|^2\sigma^{AI}_{SM}}{\sum_{SM}|c_{SM}|^2\sigma^{PI}_{SM}}-\lambda (|a_{-1}|^2+|a_{0}|^2+|a_{1}|^2-1) \right]=0,\qquad j=-1,0,1,
  \end{equation}
  \begin{equation}
  \label{eq:optimize_ratio2}
  \frac{\partial}{\partial b_j}\left[\dfrac{\sum_{SM}|c_{SM}|^2\sigma^{AI}_{SM}}{\sum_{SM}|c_{SM}|^2\sigma^{PI}_{SM}}-\gamma (|b_{-1}|^2+|b_{0}|^2+|b_{1}|^2-1) \right]=0,\qquad j=-1,0,1,
  \end{equation}
  \begin{equation}
  \label{eq:optimize_ratio3}
|a_{-1}|^2+|a_{0}|^2+|a_{1}|^2=1,
\end{equation}
\begin{equation}
\label{eq:optimize_ratio4}
|b_{-1}|^2+|b_{0}|^2+|b_{1}|^2=1.
\end{equation}
\end{subequations}
\end{widetext}
By means of the same procedure used to solve Eqs.~\eqref{eq:critical1}-\eqref{eq:critical4}, we get $\lambda=\gamma=0$. It is easy to show that the conditions~\eqref{eq:optimize_ratio1}-\eqref{eq:optimize_ratio4} are fulfilled by the coefficients~\eqref{eq:solution1}-\eqref{eq:solution3},~\ie, the ratio ${\sigma^{AI}}/{\sigma^{PI}}$ has the same critical points as the cross sections $\sigma^{AI}$ and $\sigma^{PI}$.
The numerical computations of the ionization ratio do confirm that the critical points of the ratio are the same as those of the absolute cross sections.

To search for additional  maxima (or minima) we solve Eqs.~\eqref{eq:self_consistent_a_appendix} and~\eqref{eq:self_consistent_b_appendix} [Eqs.~\eqref{eq:self_consistent_a} and~\eqref{eq:self_consistent_b} in the main text] by iteration. To explore the preparation coefficients space, we set the seed of the initial iteration by means of the coordinates $\rho, \eta, \psi, \chi,\alpha_0,\alpha_1,\beta_0$ and $\beta_1$ which define the preparation coefficients
\begin{subequations}
  \begin{equation}
    a_{0}=e^{i\alpha_0}\sin\rho\sin\eta
  \end{equation}
  \begin{equation}
    a_{1}= e^{i\alpha_1}\sin\rho\sqrt{1-\sin^2\eta}
  \end{equation}
  \begin{equation}
    a_{-1}=\sqrt{1-|a_0|^2-|a_1|^2}
  \end{equation}
  \begin{equation}
    b_{0}=e^{i\beta_0}\sin\psi\sin\chi
  \end{equation}
  \begin{equation}
    b_{1}=e^{i\beta_1}\sin\psi\sqrt{1-\sin^2\chi}
  \end{equation}
  \begin{equation}
    b_{-1}=\sqrt{1-|b_0|^2-|b_1|^2}
  \end{equation}
\end{subequations}
where  $\rho, \eta, \psi, \chi\in [0,\pi/2)$ and $\alpha_0,\alpha_1,\beta_0,\beta_1\in [0,2\pi)$. Note that $a_{-1}$ and $b_{-1}$ may be set to be real and positive to remove the redundancy on the global phase.\\


\begin{thebibliography}{44}%
\makeatletter
\providecommand \@ifxundefined [1]{%
 \@ifx{#1\undefined}
}%
\providecommand \@ifnum [1]{%
 \ifnum #1\expandafter \@firstoftwo
 \else \expandafter \@secondoftwo
 \fi
}%
\providecommand \@ifx [1]{%
 \ifx #1\expandafter \@firstoftwo
 \else \expandafter \@secondoftwo
 \fi
}%
\providecommand \natexlab [1]{#1}%
\providecommand \enquote  [1]{``#1''}%
\providecommand \bibnamefont  [1]{#1}%
\providecommand \bibfnamefont [1]{#1}%
\providecommand \citenamefont [1]{#1}%
\providecommand \href@noop [0]{\@secondoftwo}%
\providecommand \href [0]{\begingroup \@sanitize@url \@href}%
\providecommand \@href[1]{\@@startlink{#1}\@@href}%
\providecommand \@@href[1]{\endgroup#1\@@endlink}%
\providecommand \@sanitize@url [0]{\catcode `\\12\catcode `\$12\catcode
  `\&12\catcode `\#12\catcode `\^12\catcode `\_12\catcode `\%12\relax}%
\providecommand \@@startlink[1]{}%
\providecommand \@@endlink[0]{}%
\providecommand \url  [0]{\begingroup\@sanitize@url \@url }%
\providecommand \@url [1]{\endgroup\@href {#1}{\urlprefix }}%
\providecommand \urlprefix  [0]{URL }%
\providecommand \Eprint [0]{\href }%
\providecommand \doibase [0]{http://dx.doi.org/}%
\providecommand \selectlanguage [0]{\@gobble}%
\providecommand \bibinfo  [0]{\@secondoftwo}%
\providecommand \bibfield  [0]{\@secondoftwo}%
\providecommand \translation [1]{[#1]}%
\providecommand \BibitemOpen [0]{}%
\providecommand \bibitemStop [0]{}%
\providecommand \bibitemNoStop [0]{.\EOS\space}%
\providecommand \EOS [0]{\spacefactor3000\relax}%
\providecommand \BibitemShut  [1]{\csname bibitem#1\endcsname}%
\let\auto@bib@innerbib\@empty
\bibitem [{\citenamefont {Chin}\ \emph {et~al.}(2010)\citenamefont {Chin},
  \citenamefont {Grimm}, \citenamefont {Julienne},\ and\ \citenamefont
  {Tiesinga}}]{Chin:10}%
  \BibitemOpen
  \bibfield  {author} {\bibinfo {author} {\bibfnamefont {C.}~\bibnamefont
  {Chin}}, \bibinfo {author} {\bibfnamefont {R.}~\bibnamefont {Grimm}},
  \bibinfo {author} {\bibfnamefont {P.~S.}\ \bibnamefont {Julienne}}, \ and\
  \bibinfo {author} {\bibfnamefont {E.}~\bibnamefont {Tiesinga}},\ }\bibfield
  {title} {\enquote {\bibinfo {title} {Feshbach resonances in ultracold
  gases},}\ }\href {\doibase 10.1103/RevModPhys.82.1225} {\bibfield  {journal}
  {\bibinfo  {journal} {Rev. Mod. Phys.}\ }\textbf {\bibinfo {volume} {82}},\
  \bibinfo {pages} {1225--1286} (\bibinfo {year} {2010})}\BibitemShut {NoStop}%
\bibitem [{\citenamefont {Vassen}\ \emph {et~al.}(2012)\citenamefont {Vassen},
  \citenamefont {Cohen-Tannoudji}, \citenamefont {Leduc}, \citenamefont
  {Boiron}, \citenamefont {Westbrook}, \citenamefont {Truscott}, \citenamefont
  {Baldwin}, \citenamefont {Birkl}, \citenamefont {Cancio},\ and\ \citenamefont
  {Trippenbach}}]{Vassen:12}%
  \BibitemOpen
  \bibfield  {author} {\bibinfo {author} {\bibfnamefont {W.}~\bibnamefont
  {Vassen}}, \bibinfo {author} {\bibfnamefont {C.}~\bibnamefont
  {Cohen-Tannoudji}}, \bibinfo {author} {\bibfnamefont {M.}~\bibnamefont
  {Leduc}}, \bibinfo {author} {\bibfnamefont {D.}~\bibnamefont {Boiron}},
  \bibinfo {author} {\bibfnamefont {C.~I.}\ \bibnamefont {Westbrook}}, \bibinfo
  {author} {\bibfnamefont {A.}~\bibnamefont {Truscott}}, \bibinfo {author}
  {\bibfnamefont {K.}~\bibnamefont {Baldwin}}, \bibinfo {author} {\bibfnamefont
  {G.}~\bibnamefont {Birkl}}, \bibinfo {author} {\bibfnamefont
  {P.}~\bibnamefont {Cancio}}, \ and\ \bibinfo {author} {\bibfnamefont
  {M.}~\bibnamefont {Trippenbach}},\ }\bibfield  {title} {\enquote {\bibinfo
  {title} {Cold and trapped metastable noble gases},}\ }\href {\doibase
  10.1103/RevModPhys.84.175} {\bibfield  {journal} {\bibinfo  {journal} {Rev.
  Mod. Phys.}\ }\textbf {\bibinfo {volume} {84}},\ \bibinfo {pages} {175--210}
  (\bibinfo {year} {2012})}\BibitemShut {NoStop}%
\bibitem [{\citenamefont {Bloch}\ \emph {et~al.}(2008)\citenamefont {Bloch},
  \citenamefont {Dalibard},\ and\ \citenamefont {Zwerger}}]{Bloch:08}%
  \BibitemOpen
  \bibfield  {author} {\bibinfo {author} {\bibfnamefont {I.}~\bibnamefont
  {Bloch}}, \bibinfo {author} {\bibfnamefont {J.}~\bibnamefont {Dalibard}}, \
  and\ \bibinfo {author} {\bibfnamefont {W.}~\bibnamefont {Zwerger}},\
  }\bibfield  {title} {\enquote {\bibinfo {title} {Many-body physics with
  ultracold gases},}\ }\href {\doibase 10.1103/RevModPhys.80.885} {\bibfield
  {journal} {\bibinfo  {journal} {Rev. Mod. Phys.}\ }\textbf {\bibinfo {volume}
  {80}},\ \bibinfo {pages} {885--964} (\bibinfo {year} {2008})}\BibitemShut
  {NoStop}%
\bibitem [{\citenamefont {Bloch}\ \emph {et~al.}(2012)\citenamefont {Bloch},
  \citenamefont {Dalibard},\ and\ \citenamefont {Nascimb\'ene}}]{Bloch:12}%
  \BibitemOpen
  \bibfield  {author} {\bibinfo {author} {\bibfnamefont {I.}~\bibnamefont
  {Bloch}}, \bibinfo {author} {\bibfnamefont {J.}~\bibnamefont {Dalibard}}, \
  and\ \bibinfo {author} {\bibfnamefont {S.}~\bibnamefont {Nascimb\'ene}},\
  }\bibfield  {title} {\enquote {\bibinfo {title} {Quantum simulations with
  ultracold quantum gases},}\ }\href@noop {} {\bibfield  {journal} {\bibinfo
  {journal} {Nature Phys}\ }\textbf {\bibinfo {volume} {8}},\ \bibinfo {pages}
  {267--276} (\bibinfo {year} {2012})}\BibitemShut {NoStop}%
\bibitem [{\citenamefont {Krems}(2006)}]{Krems:06}%
  \BibitemOpen
  \bibfield  {author} {\bibinfo {author} {\bibfnamefont {R.~V.}\ \bibnamefont
  {Krems}},\ }\bibfield  {title} {\enquote {\bibinfo {title} {Controlling
  collisions of ultracold atoms with dc electric fields},}\ }\href {\doibase
  10.1103/PhysRevLett.96.123202} {\bibfield  {journal} {\bibinfo  {journal}
  {Phys. Rev. Lett.}\ }\textbf {\bibinfo {volume} {96}},\ \bibinfo {pages}
  {123202} (\bibinfo {year} {2006})}\BibitemShut {NoStop}%
\bibitem [{\citenamefont {Devolder}\ \emph {et~al.}(2019)\citenamefont
  {Devolder}, \citenamefont {Luc-Koenig}, \citenamefont {Atabek}, \citenamefont
  {Desouter-Lecomte},\ and\ \citenamefont {Dulieu}}]{Devolder:19}%
  \BibitemOpen
  \bibfield  {author} {\bibinfo {author} {\bibfnamefont {A.}~\bibnamefont
  {Devolder}}, \bibinfo {author} {\bibfnamefont {E.}~\bibnamefont
  {Luc-Koenig}}, \bibinfo {author} {\bibfnamefont {O.}~\bibnamefont {Atabek}},
  \bibinfo {author} {\bibfnamefont {M.}~\bibnamefont {Desouter-Lecomte}}, \
  and\ \bibinfo {author} {\bibfnamefont {O.}~\bibnamefont {Dulieu}},\
  }\bibfield  {title} {\enquote {\bibinfo {title} {Laser-assisted self-induced
  feshbach resonance for controlling heteronuclear quantum gas mixtures},}\
  }\href {\doibase 10.1103/PhysRevA.100.052703} {\bibfield  {journal} {\bibinfo
   {journal} {Phys. Rev. A}\ }\textbf {\bibinfo {volume} {100}},\ \bibinfo
  {pages} {052703} (\bibinfo {year} {2019})}\BibitemShut {NoStop}%
\bibitem [{\citenamefont {Tscherbul}\ \emph {et~al.}(2010)\citenamefont
  {Tscherbul}, \citenamefont {Calarco}, \citenamefont {Lesanovsky},
  \citenamefont {Krems}, \citenamefont {Dalgarno},\ and\ \citenamefont
  {Schmiedmayer}}]{Tscherbul:10}%
  \BibitemOpen
  \bibfield  {author} {\bibinfo {author} {\bibfnamefont {T.~V.}\ \bibnamefont
  {Tscherbul}}, \bibinfo {author} {\bibfnamefont {T.}~\bibnamefont {Calarco}},
  \bibinfo {author} {\bibfnamefont {I.}~\bibnamefont {Lesanovsky}}, \bibinfo
  {author} {\bibfnamefont {R.~V.}\ \bibnamefont {Krems}}, \bibinfo {author}
  {\bibfnamefont {A.}~\bibnamefont {Dalgarno}}, \ and\ \bibinfo {author}
  {\bibfnamefont {J.}~\bibnamefont {Schmiedmayer}},\ }\bibfield  {title}
  {\enquote {\bibinfo {title} {rf-field-induced {F}eshbach resonances},}\
  }\href {\doibase 10.1103/PhysRevA.81.050701} {\bibfield  {journal} {\bibinfo
  {journal} {Phys. Rev. A}\ }\textbf {\bibinfo {volume} {81}},\ \bibinfo
  {pages} {050701(R)} (\bibinfo {year} {2010})}\BibitemShut {NoStop}%
\bibitem [{\citenamefont {Hanna}\ \emph {et~al.}(2010)\citenamefont {Hanna},
  \citenamefont {Tiesinga},\ and\ \citenamefont {Julienne}}]{Hanna:10}%
  \BibitemOpen
  \bibfield  {author} {\bibinfo {author} {\bibfnamefont {T.~M.}\ \bibnamefont
  {Hanna}}, \bibinfo {author} {\bibfnamefont {E.}~\bibnamefont {Tiesinga}}, \
  and\ \bibinfo {author} {\bibfnamefont {P.~S.}\ \bibnamefont {Julienne}},\
  }\bibfield  {title} {\enquote {\bibinfo {title} {Creation and manipulation of
  {F}eshbach resonances with radiofrequency radiation},}\ }\href {\doibase
  10.1088/1367-2630/12/8/083031} {\bibfield  {journal} {\bibinfo  {journal}
  {New J. Phys}\ }\textbf {\bibinfo {volume} {12}},\ \bibinfo {pages} {083031}
  (\bibinfo {year} {2010})}\BibitemShut {NoStop}%
\bibitem [{\citenamefont {Ding}\ \emph {et~al.}(2017)\citenamefont {Ding},
  \citenamefont {D'Incao},\ and\ \citenamefont {Greene}}]{Ding:17}%
  \BibitemOpen
  \bibfield  {author} {\bibinfo {author} {\bibfnamefont {Y.}~\bibnamefont
  {Ding}}, \bibinfo {author} {\bibfnamefont {J.~P.}\ \bibnamefont {D'Incao}}, \
  and\ \bibinfo {author} {\bibfnamefont {C.~H.}\ \bibnamefont {Greene}},\
  }\bibfield  {title} {\enquote {\bibinfo {title} {Effective control of cold
  collisions with radio-frequency fields},}\ }\href {\doibase
  10.1103/PhysRevA.95.022709} {\bibfield  {journal} {\bibinfo  {journal} {Phys.
  Rev. A}\ }\textbf {\bibinfo {volume} {95}},\ \bibinfo {pages} {022709}
  (\bibinfo {year} {2017})}\BibitemShut {NoStop}%
\bibitem [{\citenamefont {Fedichev}\ \emph {et~al.}(1996)\citenamefont
  {Fedichev}, \citenamefont {Kagan}, \citenamefont {Shlyapnikov},\ and\
  \citenamefont {Walraven}}]{Fedichev:96}%
  \BibitemOpen
  \bibfield  {author} {\bibinfo {author} {\bibfnamefont {P.~O.}\ \bibnamefont
  {Fedichev}}, \bibinfo {author} {\bibfnamefont {Yu.}\ \bibnamefont {Kagan}},
  \bibinfo {author} {\bibfnamefont {G.~V.}\ \bibnamefont {Shlyapnikov}}, \ and\
  \bibinfo {author} {\bibfnamefont {J.~T.~M.}\ \bibnamefont {Walraven}},\
  }\bibfield  {title} {\enquote {\bibinfo {title} {Influence of nearly resonant
  light on the scattering length in low-temperature atomic gases},}\ }\href
  {\doibase 10.1103/PhysRevLett.77.2913} {\bibfield  {journal} {\bibinfo
  {journal} {Phys. Rev. Lett.}\ }\textbf {\bibinfo {volume} {77}},\ \bibinfo
  {pages} {2913--2916} (\bibinfo {year} {1996})}\BibitemShut {NoStop}%
\bibitem [{\citenamefont {Bohn}\ and\ \citenamefont
  {Julienne}(1997)}]{Bohn:97}%
  \BibitemOpen
  \bibfield  {author} {\bibinfo {author} {\bibfnamefont {J.~L.}\ \bibnamefont
  {Bohn}}\ and\ \bibinfo {author} {\bibfnamefont {P.~S.}\ \bibnamefont
  {Julienne}},\ }\bibfield  {title} {\enquote {\bibinfo {title} {Prospects for
  influencing scattering lengths with far-off-resonant light},}\ }\href
  {\doibase 10.1103/PhysRevA.56.1486} {\bibfield  {journal} {\bibinfo
  {journal} {Phys. Rev. A}\ }\textbf {\bibinfo {volume} {56}},\ \bibinfo
  {pages} {1486--1491} (\bibinfo {year} {1997})}\BibitemShut {NoStop}%
\bibitem [{\citenamefont {Nicholson}\ \emph {et~al.}(2015)\citenamefont
  {Nicholson}, \citenamefont {Blatt}, \citenamefont {Bloom}, \citenamefont
  {Williams}, \citenamefont {Thomsen}, \citenamefont {Ye},\ and\ \citenamefont
  {Julienne}}]{Nicholson:15}%
  \BibitemOpen
  \bibfield  {author} {\bibinfo {author} {\bibfnamefont {T.~L.}\ \bibnamefont
  {Nicholson}}, \bibinfo {author} {\bibfnamefont {S.}~\bibnamefont {Blatt}},
  \bibinfo {author} {\bibfnamefont {B.~J.}\ \bibnamefont {Bloom}}, \bibinfo
  {author} {\bibfnamefont {J.~R.}\ \bibnamefont {Williams}}, \bibinfo {author}
  {\bibfnamefont {J.~W.}\ \bibnamefont {Thomsen}}, \bibinfo {author}
  {\bibfnamefont {J.}~\bibnamefont {Ye}}, \ and\ \bibinfo {author}
  {\bibfnamefont {P.~S.}\ \bibnamefont {Julienne}},\ }\bibfield  {title}
  {\enquote {\bibinfo {title} {Optical feshbach resonances: Field-dressed
  theory and comparison with experiments},}\ }\href {\doibase
  10.1103/PhysRevA.92.022709} {\bibfield  {journal} {\bibinfo  {journal} {Phys.
  Rev. A}\ }\textbf {\bibinfo {volume} {92}},\ \bibinfo {pages} {022709}
  (\bibinfo {year} {2015})}\BibitemShut {NoStop}%
\bibitem [{\citenamefont {Shapiro}\ and\ \citenamefont
  {Brumer}(2012)}]{Shapiro2012}%
  \BibitemOpen
  \bibfield  {author} {\bibinfo {author} {\bibfnamefont {M.}~\bibnamefont
  {Shapiro}}\ and\ \bibinfo {author} {\bibfnamefont {P.}~\bibnamefont
  {Brumer}},\ }\href@noop {} {\emph {\bibinfo {title} {{Quantum Control of
  Molecular Processes}}}}\ (\bibinfo  {publisher} {WILEY-VCH Verlag},\ \bibinfo
  {year} {2012})\BibitemShut {NoStop}%
\bibitem [{\citenamefont {Omiste}\ \emph {et~al.}(2018)\citenamefont {Omiste},
  \citenamefont {Flo{\ss}},\ and\ \citenamefont {Brumer}}]{Omiste2018_penning}%
  \BibitemOpen
  \bibfield  {author} {\bibinfo {author} {\bibfnamefont {J.~J.}\ \bibnamefont
  {Omiste}}, \bibinfo {author} {\bibfnamefont {J.}~\bibnamefont {Flo{\ss}}}, \
  and\ \bibinfo {author} {\bibfnamefont {P.}~\bibnamefont {Brumer}},\
  }\bibfield  {title} {\enquote {\bibinfo {title} {{Coherent Control of Penning
  and Associative Ionization: Insights from Symmetries}},}\ }\href {\doibase
  https://doi.org/10.1103/PhysRevLett.121.163405} {\bibfield  {journal}
  {\bibinfo  {journal} {Phys. Rev. Lett.}\ }\textbf {\bibinfo {volume} {121}},\
  \bibinfo {pages} {163405} (\bibinfo {year} {2018})}\BibitemShut {NoStop}%
\bibitem [{\citenamefont {Devolder}\ \emph {et~al.}(2020)\citenamefont
  {Devolder}, \citenamefont {Tscherbul},\ and\ \citenamefont
  {Brumer}}]{Devolder:20}%
  \BibitemOpen
  \bibfield  {author} {\bibinfo {author} {\bibfnamefont {A.}~\bibnamefont
  {Devolder}}, \bibinfo {author} {\bibfnamefont {T.~V.}\ \bibnamefont
  {Tscherbul}}, \ and\ \bibinfo {author} {\bibfnamefont {P.}~\bibnamefont
  {Brumer}},\ }\bibfield  {title} {\enquote {\bibinfo {title} {Coherent control
  of reactive scattering at low temperatures: Signatures of quantum
  interference in the differential cross sections for {F~+~H$_2$} and
  {F~+~HD}},}\ }\href {\doibase 10.1103/PhysRevA.102.031303} {\bibfield
  {journal} {\bibinfo  {journal} {Phys. Rev. A}\ }\textbf {\bibinfo {volume}
  {102}},\ \bibinfo {pages} {031303(R)} (\bibinfo {year} {2020})}\BibitemShut
  {NoStop}%
\bibitem [{\citenamefont {Shapiro}\ and\ \citenamefont
  {Brumer}(1996)}]{Brumer1996}%
  \BibitemOpen
  \bibfield  {author} {\bibinfo {author} {\bibfnamefont {M.}~\bibnamefont
  {Shapiro}}\ and\ \bibinfo {author} {\bibfnamefont {P.}~\bibnamefont
  {Brumer}},\ }\bibfield  {title} {\enquote {\bibinfo {title} {{Coherent
  control of collisional events: Bimolecular reactive scattering}},}\ }\href
  {http://www.ncbi.nlm.nih.gov/pubmed/10061988} {\bibfield  {journal} {\bibinfo
   {journal} {Phys. Rev. Lett.}\ }\textbf {\bibinfo {volume} {77}},\ \bibinfo
  {pages} {2574} (\bibinfo {year} {1996})}\BibitemShut {NoStop}%
\bibitem [{\citenamefont {Abrashkevich}\ \emph {et~al.}(2001)\citenamefont
  {Abrashkevich}, \citenamefont {Shapiro},\ and\ \citenamefont
  {Brumer}}]{Abrashkevich2001}%
  \BibitemOpen
  \bibfield  {author} {\bibinfo {author} {\bibfnamefont {A.}~\bibnamefont
  {Abrashkevich}}, \bibinfo {author} {\bibfnamefont {M.}~\bibnamefont
  {Shapiro}}, \ and\ \bibinfo {author} {\bibfnamefont {P.}~\bibnamefont
  {Brumer}},\ }\bibfield  {title} {\enquote {\bibinfo {title} {{Coherent
  control of atom-diatom reactive scattering: Isotopic variants of H + H2 in
  three dimensions}},}\ }\href {\doibase 10.1016/S0301-0104(01)00256-7}
  {\bibfield  {journal} {\bibinfo  {journal} {Chem. Phys.}\ }\textbf {\bibinfo
  {volume} {267}},\ \bibinfo {pages} {81--92} (\bibinfo {year}
  {2001})}\BibitemShut {NoStop}%
\bibitem [{\citenamefont {Gordon}\ \emph {et~al.}(2018)\citenamefont {Gordon},
  \citenamefont {Omiste}, \citenamefont {Zou}, \citenamefont {Tanteri},
  \citenamefont {Brumer},\ and\ \citenamefont {Osterwalder}}]{Gordon2018}%
  \BibitemOpen
  \bibfield  {author} {\bibinfo {author} {\bibfnamefont {S.~D.~S.}\
  \bibnamefont {Gordon}}, \bibinfo {author} {\bibfnamefont {J.~J.}\
  \bibnamefont {Omiste}}, \bibinfo {author} {\bibfnamefont {J.}~\bibnamefont
  {Zou}}, \bibinfo {author} {\bibfnamefont {S.}~\bibnamefont {Tanteri}},
  \bibinfo {author} {\bibfnamefont {P.}~\bibnamefont {Brumer}}, \ and\ \bibinfo
  {author} {\bibfnamefont {A.}~\bibnamefont {Osterwalder}},\ }\bibfield
  {title} {\enquote {\bibinfo {title} {{Quantum-state-controlled channel
  branching in cold Ne($^3$P$_2$)+Ar chemi-ionization}},}\ }\href {\doibase
  10.1038/s41557-018-0152-2} {\bibfield  {journal} {\bibinfo  {journal} {Nat.
  Chem.}\ }\textbf {\bibinfo {volume} {10}},\ \bibinfo {pages} {1190--1195}
  (\bibinfo {year} {2018})}\BibitemShut {NoStop}%
\bibitem [{\citenamefont {Arango}\ \emph
  {et~al.}(2006{\natexlab{a}})\citenamefont {Arango}, \citenamefont {Shapiro},\
  and\ \citenamefont {Brumer}}]{Arango2006a}%
  \BibitemOpen
  \bibfield  {author} {\bibinfo {author} {\bibfnamefont {C.~A.}\ \bibnamefont
  {Arango}}, \bibinfo {author} {\bibfnamefont {M.}~\bibnamefont {Shapiro}}, \
  and\ \bibinfo {author} {\bibfnamefont {P.}~\bibnamefont {Brumer}},\
  }\bibfield  {title} {\enquote {\bibinfo {title} {{Cold atomic collisions:
  Coherent control of penning and associative ionization}},}\ }\href {\doibase
  10.1103/PhysRevLett.97.193202} {\bibfield  {journal} {\bibinfo  {journal}
  {Phys. Rev. Lett.}\ }\textbf {\bibinfo {volume} {97}},\ \bibinfo {pages}
  {193202} (\bibinfo {year} {2006}{\natexlab{a}})}\BibitemShut {NoStop}%
\bibitem [{\citenamefont {Arango}\ \emph
  {et~al.}(2006{\natexlab{b}})\citenamefont {Arango}, \citenamefont {Shapiro},\
  and\ \citenamefont {Brumer}}]{Arango2006b}%
  \BibitemOpen
  \bibfield  {author} {\bibinfo {author} {\bibfnamefont {C.~A.}\ \bibnamefont
  {Arango}}, \bibinfo {author} {\bibfnamefont {M.}~\bibnamefont {Shapiro}}, \
  and\ \bibinfo {author} {\bibfnamefont {P.}~\bibnamefont {Brumer}},\
  }\bibfield  {title} {\enquote {\bibinfo {title} {{Coherent control of
  collision processes: Penning versus associative ionization}},}\ }\href
  {\doibase 10.1063/1.2336430} {\bibfield  {journal} {\bibinfo  {journal} {J.
  Chem. Phys.}\ }\textbf {\bibinfo {volume} {125}},\ \bibinfo {pages} {094315}
  (\bibinfo {year} {2006}{\natexlab{b}})}\BibitemShut {NoStop}%
\bibitem [{Note1()}]{Note1}%
  \BibitemOpen
  \bibinfo {note} {As noted in Ref. [13], there is a computational error in
  References [17] and [18]. The formalism in these papers is correct, and
  referenced in this paper, but the computed control results are incorrect and
  superceded by Ref. [13]}\BibitemShut {NoStop}%
\bibitem [{\citenamefont {Doret}\ \emph {et~al.}(2009)\citenamefont {Doret},
  \citenamefont {Connolly}, \citenamefont {Ketterle},\ and\ \citenamefont
  {Doyle}}]{Doret:09}%
  \BibitemOpen
  \bibfield  {author} {\bibinfo {author} {\bibfnamefont {S.~C.}\ \bibnamefont
  {Doret}}, \bibinfo {author} {\bibfnamefont {C.~B.}\ \bibnamefont {Connolly}},
  \bibinfo {author} {\bibfnamefont {W.}~\bibnamefont {Ketterle}}, \ and\
  \bibinfo {author} {\bibfnamefont {J.~M.}\ \bibnamefont {Doyle}},\ }\bibfield
  {title} {\enquote {\bibinfo {title} {Buffer-gas cooled {B}ose-{E}instein
  condensate},}\ }\href {\doibase 10.1103/PhysRevLett.103.103005} {\bibfield
  {journal} {\bibinfo  {journal} {Phys. Rev. Lett.}\ }\textbf {\bibinfo
  {volume} {103}},\ \bibinfo {pages} {103005} (\bibinfo {year}
  {2009})}\BibitemShut {NoStop}%
\bibitem [{\citenamefont {{Pereira Dos Santos}}\ \emph
  {et~al.}(2001)\citenamefont {{Pereira Dos Santos}}, \citenamefont
  {L{\'{e}}onard}, \citenamefont {Wang}, \citenamefont {Barrelet},
  \citenamefont {Perales}, \citenamefont {Rasel}, \citenamefont {Unnikrishnan},
  \citenamefont {Leduc},\ and\ \citenamefont
  {Cohen-Tannoudji}}]{PereiraDosSantos2001}%
  \BibitemOpen
  \bibfield  {author} {\bibinfo {author} {\bibfnamefont {F.}~\bibnamefont
  {{Pereira Dos Santos}}}, \bibinfo {author} {\bibfnamefont {J.}~\bibnamefont
  {L{\'{e}}onard}}, \bibinfo {author} {\bibfnamefont {J.}~\bibnamefont {Wang}},
  \bibinfo {author} {\bibfnamefont {C.~J.}\ \bibnamefont {Barrelet}}, \bibinfo
  {author} {\bibfnamefont {F.}~\bibnamefont {Perales}}, \bibinfo {author}
  {\bibfnamefont {E.}~\bibnamefont {Rasel}}, \bibinfo {author} {\bibfnamefont
  {C.~S.}\ \bibnamefont {Unnikrishnan}}, \bibinfo {author} {\bibfnamefont
  {M.}~\bibnamefont {Leduc}}, \ and\ \bibinfo {author} {\bibfnamefont
  {C.}~\bibnamefont {Cohen-Tannoudji}},\ }\bibfield  {title} {\enquote
  {\bibinfo {title} {{Bose-Einstein condensation of metastable helium}},}\
  }\href {\doibase 10.1103/PhysRevLett.86.3459} {\bibfield  {journal} {\bibinfo
   {journal} {Phys. Rev. Lett.}\ }\textbf {\bibinfo {volume} {86}},\ \bibinfo
  {pages} {3459--3462} (\bibinfo {year} {2001})}\BibitemShut {NoStop}%
\bibitem [{\citenamefont {Robert}\ \emph {et~al.}(2001)\citenamefont {Robert},
  \citenamefont {Sirjean}, \citenamefont {Browaeys}, \citenamefont {Poupard},
  \citenamefont {Nowak}, \citenamefont {Boiron}, \citenamefont {Westbrook},\
  and\ \citenamefont {Aspect}}]{Robert:01}%
  \BibitemOpen
  \bibfield  {author} {\bibinfo {author} {\bibfnamefont {A.}~\bibnamefont
  {Robert}}, \bibinfo {author} {\bibfnamefont {O.}~\bibnamefont {Sirjean}},
  \bibinfo {author} {\bibfnamefont {A.}~\bibnamefont {Browaeys}}, \bibinfo
  {author} {\bibfnamefont {J.}~\bibnamefont {Poupard}}, \bibinfo {author}
  {\bibfnamefont {S.}~\bibnamefont {Nowak}}, \bibinfo {author} {\bibfnamefont
  {D.}~\bibnamefont {Boiron}}, \bibinfo {author} {\bibfnamefont {C.~I.}\
  \bibnamefont {Westbrook}}, \ and\ \bibinfo {author} {\bibfnamefont
  {A.}~\bibnamefont {Aspect}},\ }\bibfield  {title} {\enquote {\bibinfo {title}
  {A {Bose-Einstein Condensate} of metastable atoms},}\ }\href {\doibase
  10.1126/science.1060622} {\bibfield  {journal} {\bibinfo  {journal}
  {Science}\ }\textbf {\bibinfo {volume} {292}},\ \bibinfo {pages} {461}
  (\bibinfo {year} {2001})}\BibitemShut {NoStop}%
\bibitem [{\citenamefont {Barletta}\ \emph {et~al.}(2009)\citenamefont
  {Barletta}, \citenamefont {Tennyson},\ and\ \citenamefont
  {Barker}}]{Barletta2009}%
  \BibitemOpen
  \bibfield  {author} {\bibinfo {author} {\bibfnamefont {P.}~\bibnamefont
  {Barletta}}, \bibinfo {author} {\bibfnamefont {J.}~\bibnamefont {Tennyson}},
  \ and\ \bibinfo {author} {\bibfnamefont {P.~F.}\ \bibnamefont {Barker}},\
  }\bibfield  {title} {\enquote {\bibinfo {title} {{Towards sympathetic cooling
  of large molecules: cold collisions between benzene and rare gas atoms}},}\
  }\href {\doibase 10.1088/1367-2630/11/5/055029} {\bibfield  {journal}
  {\bibinfo  {journal} {New J. Phys.}\ }\textbf {\bibinfo {volume} {11}},\
  \bibinfo {pages} {55029} (\bibinfo {year} {2009})}\BibitemShut {NoStop}%
\bibitem [{\citenamefont {Tscherbul}\ \emph {et~al.}(2011)\citenamefont
  {Tscherbul}, \citenamefont {Yu},\ and\ \citenamefont
  {Dalgarno}}]{Tscherbul2011}%
  \BibitemOpen
  \bibfield  {author} {\bibinfo {author} {\bibfnamefont {T.~V.}\ \bibnamefont
  {Tscherbul}}, \bibinfo {author} {\bibfnamefont {H.-G.}\ \bibnamefont {Yu}}, \
  and\ \bibinfo {author} {\bibfnamefont {A.}~\bibnamefont {Dalgarno}},\
  }\bibfield  {title} {\enquote {\bibinfo {title} {{Sympathetic Cooling of
  Polyatomic Molecules with $S$-State Atoms in a Magnetic Trap}},}\ }\href
  {\doibase 10.1103/PhysRevLett.106.073201} {\bibfield  {journal} {\bibinfo
  {journal} {Phys. Rev. Lett.}\ }\textbf {\bibinfo {volume} {106}},\ \bibinfo
  {pages} {073201} (\bibinfo {year} {2011})}\BibitemShut {NoStop}%
\bibitem [{\citenamefont {Knoop}\ \emph {et~al.}(2014)\citenamefont {Knoop},
  \citenamefont {{\.Z}uchowski}, \citenamefont {Kedziera}, \citenamefont
  {Mentel}, \citenamefont {Puchalski}, \citenamefont {Mishra}, \citenamefont
  {Flores},\ and\ \citenamefont {Vassen}}]{Knoop:14}%
  \BibitemOpen
  \bibfield  {author} {\bibinfo {author} {\bibfnamefont {S.}~\bibnamefont
  {Knoop}}, \bibinfo {author} {\bibfnamefont {P.S.}\ \bibnamefont
  {{\.Z}uchowski}}, \bibinfo {author} {\bibfnamefont {D.}~\bibnamefont
  {Kedziera}}, \bibinfo {author} {\bibfnamefont {\L.}\ \bibnamefont {Mentel}},
  \bibinfo {author} {\bibfnamefont {M.}~\bibnamefont {Puchalski}}, \bibinfo
  {author} {\bibfnamefont {H.~P.}\ \bibnamefont {Mishra}}, \bibinfo {author}
  {\bibfnamefont {A.~S.}\ \bibnamefont {Flores}}, \ and\ \bibinfo {author}
  {\bibfnamefont {W.}~\bibnamefont {Vassen}},\ }\bibfield  {title} {\enquote
  {\bibinfo {title} {{Ultracold mixtures of metastable He and Rb: Scattering
  lengths from ab initio calculations and thermalization measurements}},}\
  }\href {\doibase 10.1103/PhysRevA.90.022709} {\bibfield  {journal} {\bibinfo
  {journal} {Phys. Rev. A}\ }\textbf {\bibinfo {volume} {90}},\ \bibinfo
  {pages} {22709} (\bibinfo {year} {2014})}\BibitemShut {NoStop}%
\bibitem [{\citenamefont {Siska}(1993)}]{Siska1993}%
  \BibitemOpen
  \bibfield  {author} {\bibinfo {author} {\bibfnamefont {P.~E.}\ \bibnamefont
  {Siska}},\ }\bibfield  {title} {\enquote {\bibinfo {title} {{Molecular-beam
  studies of Penning ionization}},}\ }\href {\doibase
  10.1103/RevModPhys.65.337} {\bibfield  {journal} {\bibinfo  {journal} {Rev.
  Mod. Phys.}\ }\textbf {\bibinfo {volume} {65}},\ \bibinfo {pages} {337--412}
  (\bibinfo {year} {1993})}\BibitemShut {NoStop}%
\bibitem [{\citenamefont {M{\"{u}}ller}\ \emph {et~al.}(1991)\citenamefont
  {M{\"{u}}ller}, \citenamefont {Merz}, \citenamefont {Ruf}, \citenamefont
  {Hotop}, \citenamefont {Meyer},\ and\ \citenamefont {Movre}}]{Muller1991}%
  \BibitemOpen
  \bibfield  {author} {\bibinfo {author} {\bibfnamefont {M.~W.}\ \bibnamefont
  {M{\"{u}}ller}}, \bibinfo {author} {\bibfnamefont {A.}~\bibnamefont {Merz}},
  \bibinfo {author} {\bibfnamefont {M.~W.~W.}\ \bibnamefont {Ruf}}, \bibinfo
  {author} {\bibfnamefont {H.}~\bibnamefont {Hotop}}, \bibinfo {author}
  {\bibfnamefont {W.}~\bibnamefont {Meyer}}, \ and\ \bibinfo {author}
  {\bibfnamefont {M.}~\bibnamefont {Movre}},\ }\bibfield  {title} {\enquote
  {\bibinfo {title} {{Experimental and theoretical studies of the Bi-excited
  collision systems He$^*(2{}^3\text{S})$ + He$^*(2{}^3\text{S},
  \text{2}{}^1\text{S})$ at thermal and subthermal kinetic energies}},}\ }\href
  {\doibase 10.1007/BF01425589} {\bibfield  {journal} {\bibinfo  {journal}
  {Zeitschrift f{{\"{u}}}r Phys. D Atoms, Mol. Clust.}\ }\textbf {\bibinfo
  {volume} {21}},\ \bibinfo {pages} {89--112} (\bibinfo {year}
  {1991})}\BibitemShut {NoStop}%
\bibitem [{\citenamefont {Kramida}\ \emph {et~al.}(2018)\citenamefont
  {Kramida}, \citenamefont {Ralchenko},\ and\ \citenamefont
  {Reader}}]{NIST_ASD_2019}%
  \BibitemOpen
  \bibfield  {author} {\bibinfo {author} {\bibfnamefont {A.}~\bibnamefont
  {Kramida}}, \bibinfo {author} {\bibfnamefont {Yu.}\ \bibnamefont
  {Ralchenko}}, \ and\ \bibinfo {author} {\bibfnamefont {J.}~\bibnamefont
  {Reader}},\ }\href@noop {} {}\bibinfo {howpublished} {NIST Atomic Spectra
  Database (ver. 5.6.1), [Online]. Available:
  {\tt{https://physics.nist.gov/asd}}. National Institute of Standards and
  Technology, Gaithersburg, MD.} (\bibinfo {year} {2018})\BibitemShut {NoStop}%
\bibitem [{\citenamefont {Mori}\ \emph {et~al.}(1964)\citenamefont {Mori},
  \citenamefont {Watanabe},\ and\ \citenamefont {Katsuura}}]{Mori1964}%
  \BibitemOpen
  \bibfield  {author} {\bibinfo {author} {\bibfnamefont {M.}~\bibnamefont
  {Mori}}, \bibinfo {author} {\bibfnamefont {T.}~\bibnamefont {Watanabe}}, \
  and\ \bibinfo {author} {\bibfnamefont {K.}~\bibnamefont {Katsuura}},\
  }\bibfield  {title} {\enquote {\bibinfo {title} {{Collisional excitation
  transfer between atoms in the resonant process}},}\ }\href {\doibase
  10.1143/JPSJ.19.380} {\bibfield  {journal} {\bibinfo  {journal} {J. Phys.
  Soc. Japan}\ }\textbf {\bibinfo {volume} {19}},\ \bibinfo {pages} {380--386}
  (\bibinfo {year} {1964})}\BibitemShut {NoStop}%
\bibitem [{\citenamefont {Hill}\ \emph {et~al.}(1972)\citenamefont {Hill},
  \citenamefont {Hatfield}, \citenamefont {Stockwell},\ and\ \citenamefont
  {Walters}}]{Hill1972}%
  \BibitemOpen
  \bibfield  {author} {\bibinfo {author} {\bibfnamefont {J.~C.}\ \bibnamefont
  {Hill}}, \bibinfo {author} {\bibfnamefont {L.~L.}\ \bibnamefont {Hatfield}},
  \bibinfo {author} {\bibfnamefont {N.~D.}\ \bibnamefont {Stockwell}}, \ and\
  \bibinfo {author} {\bibfnamefont {G.~K.}\ \bibnamefont {Walters}},\
  }\bibfield  {title} {\enquote {\bibinfo {title} {{Direct Demonstration of
  Spin-Angular-Momentum Conservation in the Reaction
  $\mathrm{He}(2^{2}S_{1})+\mathrm{He}(2^{3}S_{1})\ensuremath{\rightarrow}\mathrm{He}(1^{1}S_{0})+{\mathrm{He}}^{+}+{e}^{\ensuremath{-}}$}},}\
  }\href {\doibase 10.1103/PhysRevA.5.189} {\bibfield  {journal} {\bibinfo
  {journal} {Phys. Rev. A}\ }\textbf {\bibinfo {volume} {5}},\ \bibinfo {pages}
  {189--195} (\bibinfo {year} {1972})}\BibitemShut {NoStop}%
\bibitem [{\citenamefont {Garrison}\ \emph {et~al.}(1973)\citenamefont
  {Garrison}, \citenamefont {Miller},\ and\ \citenamefont
  {Schaefer}}]{Garrison1973}%
  \BibitemOpen
  \bibfield  {author} {\bibinfo {author} {\bibfnamefont {B.~J.}\ \bibnamefont
  {Garrison}}, \bibinfo {author} {\bibfnamefont {W.~H.}\ \bibnamefont
  {Miller}}, \ and\ \bibinfo {author} {\bibfnamefont {H.~F.}\ \bibnamefont
  {Schaefer}},\ }\bibfield  {title} {\enquote {\bibinfo {title} {{Penning and
  associative ionization of triplet metastable helium atoms}},}\ }\href
  {\doibase 10.1063/1.1680460} {\bibfield  {journal} {\bibinfo  {journal} {J.
  Chem. Phys.}\ }\textbf {\bibinfo {volume} {59}},\ \bibinfo {pages} {3193}
  (\bibinfo {year} {1973})}\BibitemShut {NoStop}%
\bibitem [{\citenamefont {M{\"{u}}ller}\ \emph {et~al.}(1987)\citenamefont
  {M{\"{u}}ller}, \citenamefont {Bussert}, \citenamefont {Ruf}, \citenamefont
  {Hotop},\ and\ \citenamefont {Meyer}}]{Muller1987}%
  \BibitemOpen
  \bibfield  {author} {\bibinfo {author} {\bibfnamefont {M.~W.}\ \bibnamefont
  {M{\"{u}}ller}}, \bibinfo {author} {\bibfnamefont {W.}~\bibnamefont
  {Bussert}}, \bibinfo {author} {\bibfnamefont {M.-W.}\ \bibnamefont {Ruf}},
  \bibinfo {author} {\bibfnamefont {H.}~\bibnamefont {Hotop}}, \ and\ \bibinfo
  {author} {\bibfnamefont {W.}~\bibnamefont {Meyer}},\ }\bibfield  {title}
  {\enquote {\bibinfo {title} {{New oscillatory structure in electron energy
  spectra from autoionizing quasi-molecules: Subthermal collisions of
  He${(2}^{3}$S) atoms with He${(2}^{1}$S,${2}^{3}$S) atoms}},}\ }\href
  {\doibase 10.1103/PhysRevLett.59.2279} {\bibfield  {journal} {\bibinfo
  {journal} {Phys. Rev. Lett.}\ }\textbf {\bibinfo {volume} {59}},\ \bibinfo
  {pages} {2279--2282} (\bibinfo {year} {1987})}\BibitemShut {NoStop}%
\bibitem [{\citenamefont {Leo}\ \emph {et~al.}(2001)\citenamefont {Leo},
  \citenamefont {Venturi}, \citenamefont {Whittingham},\ and\ \citenamefont
  {Babb}}]{Leo2001}%
  \BibitemOpen
  \bibfield  {author} {\bibinfo {author} {\bibfnamefont {P.~J.}\ \bibnamefont
  {Leo}}, \bibinfo {author} {\bibfnamefont {V.}~\bibnamefont {Venturi}},
  \bibinfo {author} {\bibfnamefont {I.~B.}\ \bibnamefont {Whittingham}}, \ and\
  \bibinfo {author} {\bibfnamefont {J.~F.}\ \bibnamefont {Babb}},\ }\bibfield
  {title} {\enquote {\bibinfo {title} {{Ultracold collisions of metastable
  helium atoms}},}\ }\href {\doibase 10.1103/PhysRevA.64.042710} {\bibfield
  {journal} {\bibinfo  {journal} {Phys. Rev. A}\ }\textbf {\bibinfo {volume}
  {64}},\ \bibinfo {pages} {042710} (\bibinfo {year}
  {2001})}\BibitemShut {NoStop}%
\bibitem [{\citenamefont {Stas}\ \emph {et~al.}(2006)\citenamefont {Stas},
  \citenamefont {McNamara}, \citenamefont {Hogervorst},\ and\ \citenamefont
  {Vassen}}]{Stas2006}%
  \BibitemOpen
  \bibfield  {author} {\bibinfo {author} {\bibfnamefont {R.~J.~W.}\
  \bibnamefont {Stas}}, \bibinfo {author} {\bibfnamefont {J.~M.}\ \bibnamefont
  {McNamara}}, \bibinfo {author} {\bibfnamefont {W.}~\bibnamefont
  {Hogervorst}}, \ and\ \bibinfo {author} {\bibfnamefont {W.}~\bibnamefont
  {Vassen}},\ }\bibfield  {title} {\enquote {\bibinfo {title} {{Homonuclear
  ionizing collisions of laser-cooled metastable helium atoms}},}\ }\href
  {\doibase 10.1103/PhysRevA.73.032713} {\bibfield  {journal} {\bibinfo
  {journal} {Phys. Rev. A}\ }\textbf {\bibinfo {volume} {73}},\ \bibinfo
  {pages} {032713} (\bibinfo {year} {2006})}\BibitemShut {NoStop}%
\bibitem [{\citenamefont {Xie}\ \emph {et~al.}(2005)\citenamefont {Xie},
  \citenamefont {Poirier},\ and\ \citenamefont {Gellene}}]{Xie2005}%
  \BibitemOpen
  \bibfield  {author} {\bibinfo {author} {\bibfnamefont {J.}~\bibnamefont
  {Xie}}, \bibinfo {author} {\bibfnamefont {B.}~\bibnamefont {Poirier}}, \ and\
  \bibinfo {author} {\bibfnamefont {G.~I.}\ \bibnamefont {Gellene}},\
  }\bibfield  {title} {\enquote {\bibinfo {title} {{Accurate, two-state ab
  initio study of the ground and first-excited states of He$_2^+$, including
  exact treatment of all Born-Oppenheimer correction terms}},}\ }\href
  {\doibase 10.1063/1.1891685} {\bibfield  {journal} {\bibinfo  {journal} {J.
  Chem. Phys.}\ }\textbf {\bibinfo {volume} {122}},\ \bibinfo {pages} {184310}
  (\bibinfo {year} {2005})}\BibitemShut {NoStop}%
\bibitem [{\citenamefont {Zou}\ \emph {et~al.}(2018)\citenamefont {Zou},
  \citenamefont {Gordon}, \citenamefont {Tanteri},\ and\ \citenamefont
  {Osterwalder}}]{Zou2018}%
  \BibitemOpen
  \bibfield  {author} {\bibinfo {author} {\bibfnamefont {J.}~\bibnamefont
  {Zou}}, \bibinfo {author} {\bibfnamefont {S.~D.~S.}\ \bibnamefont {Gordon}},
  \bibinfo {author} {\bibfnamefont {S.}~\bibnamefont {Tanteri}}, \ and\
  \bibinfo {author} {\bibfnamefont {A.}~\bibnamefont {Osterwalder}},\
  }\bibfield  {title} {\enquote {\bibinfo {title} {{Stereodynamics of
  Ne($^3\text{P}_2$) reacting with Ar, Kr, Xe, and N$_2$}},}\ }\href {\doibase
  10.1063/1.5026952} {\bibfield  {journal} {\bibinfo  {journal} {J. Chem.
  Phys.}\ }\textbf {\bibinfo {volume} {148}},\ \bibinfo {pages} {164310}
  (\bibinfo {year} {2018})}\BibitemShut {NoStop}%
\bibitem [{\citenamefont {Zare}(1988)}]{Zare1988}%
  \BibitemOpen
  \bibfield  {author} {\bibinfo {author} {\bibfnamefont {R.~N.}\ \bibnamefont
  {Zare}},\ }\href@noop {} {\emph {\bibinfo {title} {{Angular Momentum:
  Understanding Spatial Aspects in Chemistry and Physics}}}}\ (\bibinfo
  {publisher} {John Wiley and Sons},\ \bibinfo {address} {New York},\ \bibinfo
  {year} {1988})\BibitemShut {NoStop}%
\bibitem [{\citenamefont {Arimondo}(1996)}]{Arimondo1996}%
  \BibitemOpen
  \bibfield  {author} {\bibinfo {author} {\bibfnamefont {E.}~\bibnamefont
  {Arimondo}},\ }\bibfield  {title} {\enquote {\bibinfo {title} {{V Coherent
  Population Trapping in Laser Spectroscopy}},}\ \ }(\bibinfo  {publisher}
  {Elsevier},\ \bibinfo {year} {1996})\ pp.\ \bibinfo {pages}
  {257--354}\BibitemShut {NoStop}%
\bibitem [{\citenamefont {Fleischhauer}\ \emph {et~al.}(2005)\citenamefont
  {Fleischhauer}, \citenamefont {Imamoglu},\ and\ \citenamefont
  {Marangos}}]{Fleischhauer2005}%
  \BibitemOpen
  \bibfield  {author} {\bibinfo {author} {\bibfnamefont {M.}~\bibnamefont
  {Fleischhauer}}, \bibinfo {author} {\bibfnamefont {A.}~\bibnamefont
  {Imamoglu}}, \ and\ \bibinfo {author} {\bibfnamefont {J.~P.}\ \bibnamefont
  {Marangos}},\ }\bibfield  {title} {\enquote {\bibinfo {title}
  {{Electromagnetically induced transparency: Optics in coherent media}},}\
  }\href {\doibase 10.1103/RevModPhys.77.633} {\bibfield  {journal} {\bibinfo
  {journal} {Rev. Mod. Phys.}\ }\textbf {\bibinfo {volume} {77}},\ \bibinfo
  {pages} {633--673} (\bibinfo {year} {2005})}\BibitemShut {NoStop}%
\bibitem [{\citenamefont {Vitanov}\ \emph {et~al.}(2017)\citenamefont
  {Vitanov}, \citenamefont {Rangelov}, \citenamefont {Shore},\ and\
  \citenamefont {Bergmann}}]{Vitanov2017}%
  \BibitemOpen
  \bibfield  {author} {\bibinfo {author} {\bibfnamefont {N.~V.}\ \bibnamefont
  {Vitanov}}, \bibinfo {author} {\bibfnamefont {A.~A.}\ \bibnamefont
  {Rangelov}}, \bibinfo {author} {\bibfnamefont {B.~W.}\ \bibnamefont {Shore}},
  \ and\ \bibinfo {author} {\bibfnamefont {K.}~\bibnamefont {Bergmann}},\
  }\bibfield  {title} {\enquote {\bibinfo {title} {{Stimulated Raman adiabatic
  passage in physics, chemistry, and beyond}},}\ }\href {\doibase
  10.1103/RevModPhys.89.015006} {\bibfield  {journal} {\bibinfo  {journal}
  {Rev. Mod. Phys.}\ }\textbf {\bibinfo {volume} {89}},\ \bibinfo {pages}
  {015006} (\bibinfo {year} {2017})}\BibitemShut {NoStop}%
\bibitem [{\citenamefont {Taylor}(1972)}]{Taylor1972}%
  \BibitemOpen
  \bibfield  {author} {\bibinfo {author} {\bibfnamefont {J.~R.}\ \bibnamefont
  {Taylor}},\ }\href@noop {} {\emph {\bibinfo {title} {{Scattering Theory: The
  Quantum Theory of Nonrelativistic Collisions}}}}\ (\bibinfo  {publisher} {New
  York Wiley},\ \bibinfo {year} {1972})\BibitemShut {NoStop}%
\bibitem [{\citenamefont {Khan}\ \emph {et~al.}(1991)\citenamefont {Khan},
  \citenamefont {Siddiqui},\ and\ \citenamefont {Siska}}]{Khan1991}%
  \BibitemOpen
  \bibfield  {author} {\bibinfo {author} {\bibfnamefont {A.}~\bibnamefont
  {Khan}}, \bibinfo {author} {\bibfnamefont {H.~R.}\ \bibnamefont {Siddiqui}},
  \ and\ \bibinfo {author} {\bibfnamefont {P.~E.}\ \bibnamefont {Siska}},\
  }\bibfield  {title} {\enquote {\bibinfo {title} {{Angle-energy distributions
  of Penning ions in crossed molecular beams. II. Effect of Penning electron
  recoil in Ne$^*(3 s {}^3P_2)+\text{H,D}\rightarrow \text{Ne}+\text{H}^+
  ,\text{D}^+ + e^-$}},}\ }\href {\doibase 10.1063/1.460842} {\bibfield
  {journal} {\bibinfo  {journal} {J. Chem. Phys.}\ }\textbf {\bibinfo {volume}
  {95}},\ \bibinfo {pages} {3371--3380} (\bibinfo {year} {1991})}\BibitemShut
  {NoStop}%
\end{thebibliography}
\end{document}